\documentclass[amssymb,aps,showpacs,showkeys]{revtex4}

\usepackage{dcolumn}
\usepackage{graphicx}
\usepackage{bm}

\begin{document}

\title[Modes of von-Karman Covariance]{Modal Decomposition of the von-K\'arm\'an Covariance of Atmospheric Turbulence in the Circular Entrance Pupil}

\author{Richard J. Mathar}
\homepage{http://www.strw.leidenuniv.nl/~mathar}
\email{mathar@strw.leidenuniv.nl}
\affiliation{Leiden Observatory, P.O. Box 9513, 2300 RA Leiden, The Netherlands}


\date{\today}

\begin{abstract}
Estimators of outer scales of atmospheric turbulence usually fit the phase
screen snapshots derived from local wave front sensors to a Zernike basis,
and then compare the spectrum of expansion coefficients in this basis with a
narrowing
associated
with decreasing outer scales.

This manuscript discusses aspects that arise if the Zernike basis is exchanged for
a Karhunen-Lo\`eve basis of statistically independent modes.
Data acquisition turns out to be more demanding because
sensing the tip-tilt mode \emph{is} required. The data reduction
methodology
is replaced by
fitting of variance ratios of an entire
(long exposure) data set.
Statistical testing of hypotheses on outer scale models can
be applied to a set of modes---supposing other noise originating from 
detector readout and the optical train can be disentangled.

\end{abstract}

\maketitle
\section{Statistics of Atmospheric Turbulence} \label{sec.intro} 

\subsection{Measuring and Interpreting Pupil Phases}
A contemporary adaptive optics (AO) system of an astronomic telescope measures
angles of arrival of the incoming wave front sampled across
segmented areas that subdivide the full entrance pupil. The task is to weave
the two principal parameters---motion of the image either represented
by two Cartesian directions on the detector or by an angle of arrival and
azimuth \cite{GlindemannJOSAA11}---of the sensor array
into a phase function
covering the entire pupil.
For the purpose of the present work we leave aside aspects
of phase unwrapping \cite{MearaJOSA67,LloydHartJOSAA11,FriedJOSA67},
finite element discretization \cite{PearsonJOSA67},
or noise besides the atmospheric turbulence itself \cite{BechetJOSAA26}.

As the information obtained represents the phase in the pupil,
optical path differences integrated along paths through the
entire
atmosphere and altitudes dependencies are basically not available within a single AO system.
Thicknesses and altitude variation of the structure constants
may vary \cite{CoulmanAO30}.
Understanding of the origin of finite outer scales of turbulence
remains poor for that reason, and
the scattering of outer scale lengths at wavelengths in the visible is significant
\cite{AvilaJOSAA14,ConanSPIE2000,TakatoJOSAA12,CoulmanAO27}.

Optical path differences from stellar interferometers with baseline lengths
of the order of the outer scales promise higher sensitivity. If the
interferometric subsystems do not share any data streams with the 
AO systems of the individual telescopes, the time series
of the fringe tracker may be the only information available, and then
mediation by some wind velocity with its own set of unknowns
is needed to fit the shape of temporal spectra \cite{CliffordJOSA61,ColavitaAO26_4106,BuscherAO34}
to outer scales. Results from the Palomar Testbed Interferometer
merged with tip-tilt information of the telescopes show outer scales
of some tens of meters---compatible with data from single-telescope apertures \cite{LinfieldApJ554}.

Auxiliary data products of this kind are
not a standard
of reduction pipelines
of interferometric observations for now, so we discuss only the case
of a single circular aperture.
The remainder of Section \ref{sec.intro} summarizes
the literature on the von K\'arm\'an model of outer scales
with a single layer with isotropic turbulence.
Section \ref{sec.2d}
introduces its modal composition for the circular pupil. On the technical
side, the matrix elements
that arise in the Zernike basis are reduced to Hypergeometric Functions (Section \ref{sec.p2d}),
and
central obscurations common to on-axis mirror trains are handled
in a hybrid form that separates Fourier and real space with the aid of
a Neuman series (Section \ref{sec.obsc}).
Benefits and problems compared to the use of Zernike covariances
are shortly discussed in Section \ref{sec.meth}.

\subsection{The von K\'arm\'an Model of Outer Scales}

The refractive index structure function ${\cal D}_n$ as a function of the
3D
separation
$\Delta {\bf r}$, and its Fourier representation $\Phi_n$ at spatial frequencies
$f$ are related via
\begin{equation}
{\cal D}_n(\Delta {\bf r}) = 2\int \Phi_n(f)[1-\cos(2\pi{\bf f}\cdot \Delta {\bf r})]d^3f.
\end{equation}
The von K\'arm\'an model of the power spectrum in particular is
\begin{equation}
\Phi_n(f)
= c_n C_n^2(f^2+f_0^2)^{-(\gamma+3)/2}
\label{eq.vK}
\end{equation}
where $C_n^2$ is the structure constant which characterizes the global
strength of the turbulence, and where $f_0$ is a wavenumber equal to the inverse
of the outer scale. $\gamma$ is a power spectral index which for
our analysis will be set to the Kolmogorov value of $2/3$;
formulas will generally not be reduced with this value to support independent
tests of this scaling \cite{WaltersJOSA69,LewisJOSA67,BuserJOSA61,DaytonOL17,BoremanJOSAA13}.
The constant
\begin{equation}
c_n
= -\frac{\Gamma(\frac{3+\gamma}{2})}{2\pi^{3/2+\gamma}\Gamma(\frac{-\gamma}{2})}
\approx 0.00969315\quad (\gamma=2/3)
\label{eq.c3d}
\end{equation}
is
settled
by demanding that at small $\Delta {\bf r}$ 
\begin{equation}
{\cal D}_n(\Delta r) = C_n^2 |\Delta r|^\gamma.
\end{equation}

\begin{figure}[hbt]
\includegraphics[scale=0.5]{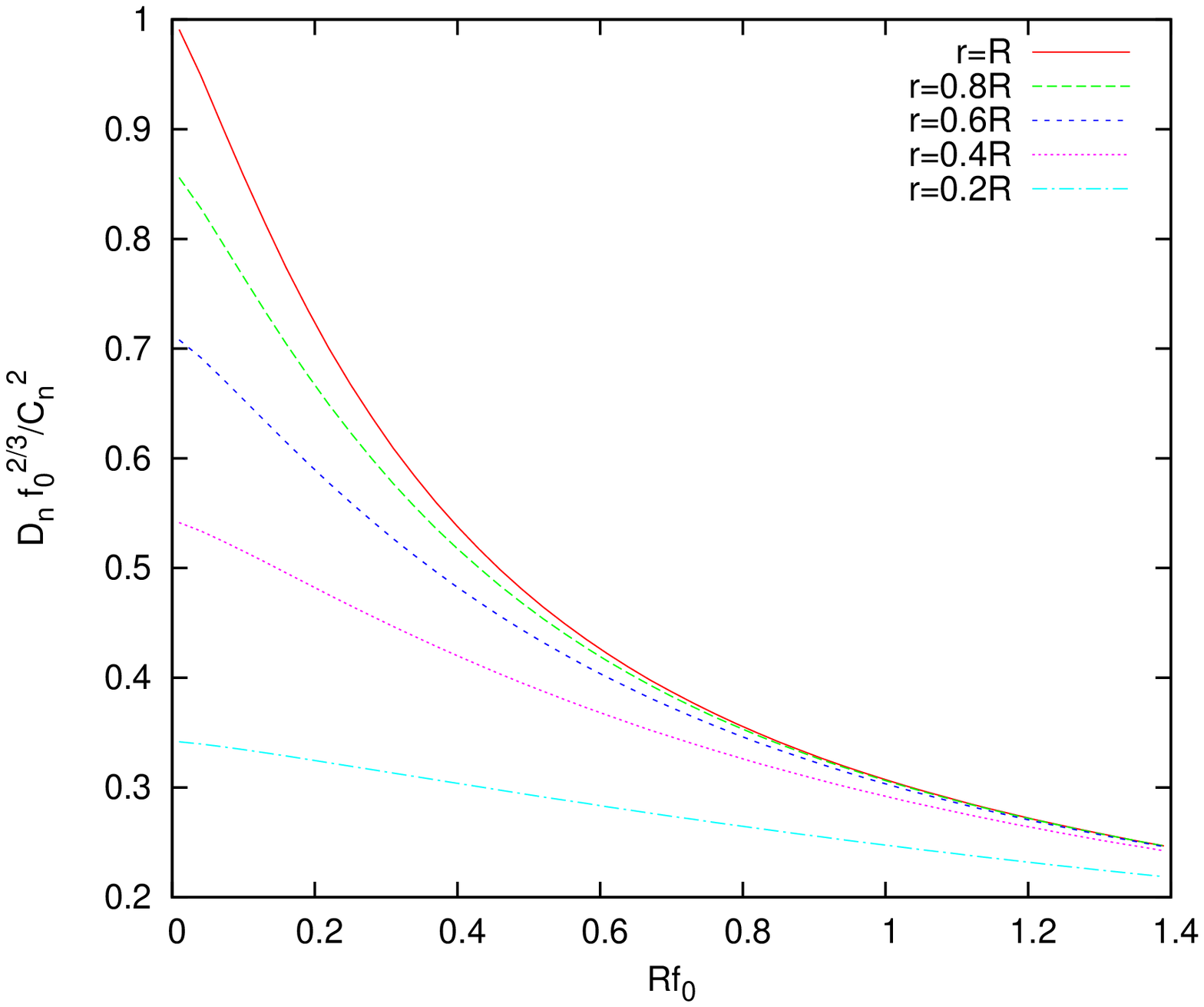}
\includegraphics[scale=0.5]{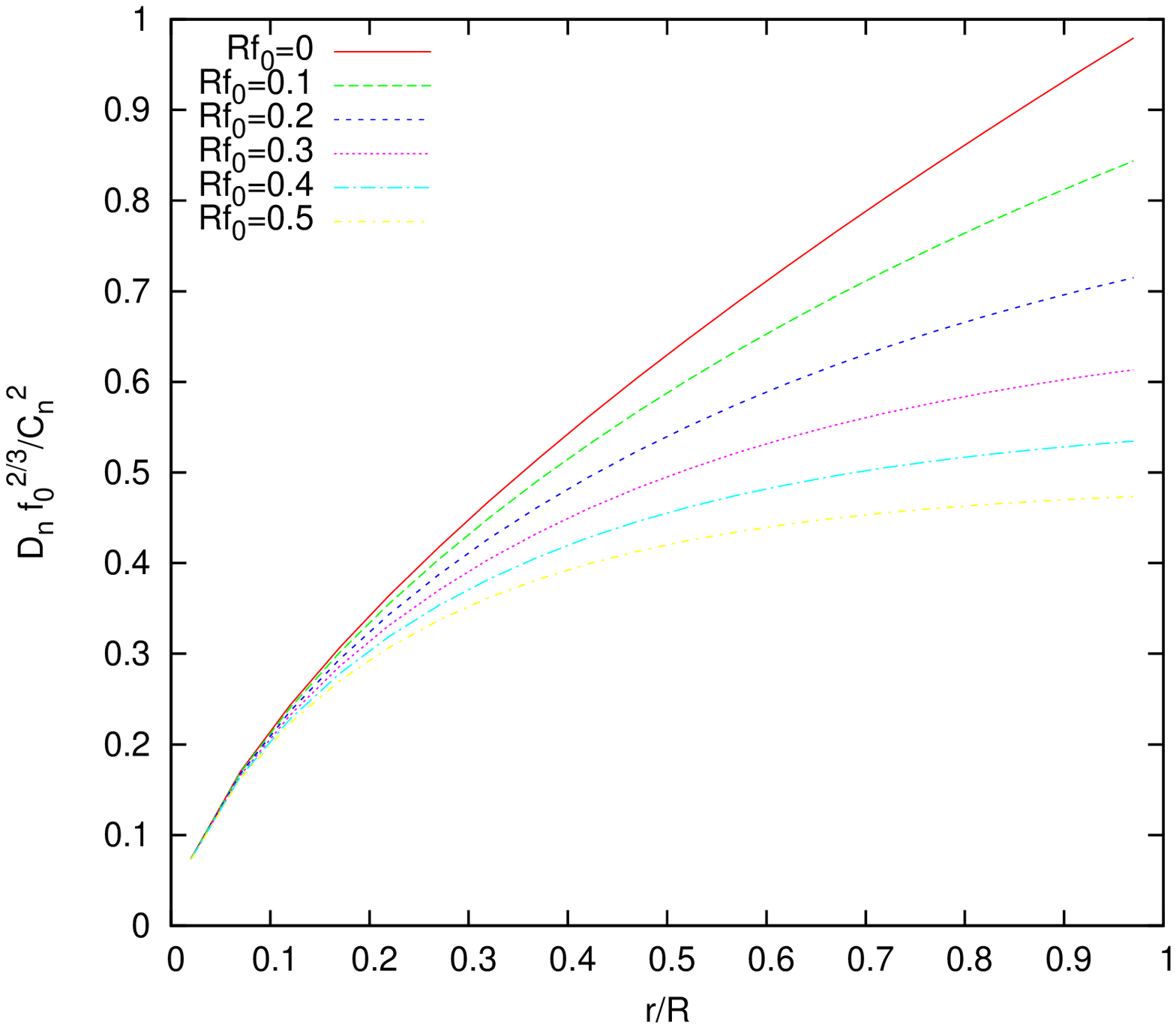}
\caption{The refractive index structure function ${\cal D}_n(r)/C_n^2$ 
of (\ref{eq.Dn}) at $\gamma=2/3$ as a function of outer scale
with the radial distance $r$ as the parameter, or with the roles of these two
dependencies swapped. ${\cal D}_n$ depends essentially on the product $rf_0$,
so scaling with some anonymous reference length $R$ leads to the universal
curves shown.
}
\label{fig.tst}
\end{figure}

This is a model of isotropic turbulence, so the information can be
bound to the radial dependencies in real and Fourier space by
integration over polar and azimuth angles,
\begin{equation}
{\cal D}_n(r)= 8\pi\int_0^\infty f^2[1-j_0(2\pi fr)]\Phi_n(f)df
= 8\pi c_n C_n^2\int_0^\infty f^2[1-j_0(2\pi fr)]\frac{1}{(f^2+f_0^2)^{(3+\gamma)/2}}df
\label{eq.Dnr}
.
\end{equation}
Performing the integration leads to a representation in terms of Modified
Bessel Functions $K$ \cite{VoitsekhoJOSAA12a,ConanJOSAA25,Conan00},
\begin{equation}
\frac{1}{C_n^2}{\cal D}_n(r) r^{-\gamma}
=
-\frac{(\pi rf_0)^{-\gamma}\Gamma(\gamma/2)}{\Gamma(-\gamma/2)}
\left[
1 - \frac{2}{\Gamma(\gamma/2)} (\pi rf_0)^{\gamma/2}K_{\gamma/2}(2\pi rf_0)
\right]
.
\label{eq.Dn}
\end{equation}
The limit of infinite outer scale recovers the Kolmogorov power law
\begin{equation}
\lim_{f_0\to 0} \frac{1}{C_n^2}{\cal D}_n(r) r^{-\gamma}
= 1,
\end{equation}
which is the value reached by the curve of $r=R$ in the left graph of Fig.\ \ref{fig.tst}.

\subsection{Phase Structure Functions}

In most applications, the observable is not the three-dimensional
distribution of the refractive indices but the phase difference $\varphi$
between light rays at an optical wavenumber $\bar k$ (which is $2\pi$ divided
by the wavelength). We write ${\cal D}_\varphi$ for 
the structure function of the phase if observed over a projected distance $P$ in
an input pupil of the instrument after two rays
have traveled along parallel lines through a layer of height $h$ (which includes
the air mass factor) in which the refractive index structure function modeled
in the previous section defines the statistics.
This type of projection onto a two-dimensional plane introduces
a weighting of the refractive index power spectrum $\Phi_n$ with Bessel Functions $J_0$
\cite{Goodman,ChesnokovOC141,ConanJOSAA25,EllerbroekAO36},
\begin{equation}
{\cal D}_\varphi(P)
=4\pi \bar k^2 h\int_0^\infty [1-J_0(2\pi fP)]\Phi_n(f)f df,
\label{eq.Dphase}
\end{equation}
for the von K\'arm\'an spectrum (\ref{eq.vK}) condensed into the well-known
\begin{equation}
{\cal D}_\varphi(P)
=
- \bar k^2 h C_n^2 P^{1+\gamma}
\frac{\sqrt{\pi}}{\Gamma(-\gamma/2)(\pi Pf_0)^{1+\gamma}}
\left[\Gamma\left(\frac{1+\gamma}{2}\right)
- 2(\pi Pf_0)^{(1+\gamma)/2}K_{(1+\gamma)/2}(2\pi Pf_0)\right]
.
\end{equation}
The Kolmogorov limit of infinite outer scale is the familiar
\begin{equation}
\lim_{f_0\to 0}
{\cal D}_\varphi(P)
=
\bar k^2 h C_n^2 P^{1+\gamma} g(\gamma)
\label{eq.dKlim}
\end{equation}
where
\begin{equation}
g(\gamma)\equiv \sqrt{\pi} \frac{\Gamma(-1/2-\gamma/2)}{\Gamma(-\gamma/2)}
\approx 2.9143808, \quad (\gamma=2/3).
\label{eq.gofgamm}
\end{equation}

\begin{figure}[hbt]
\includegraphics[scale=0.5]{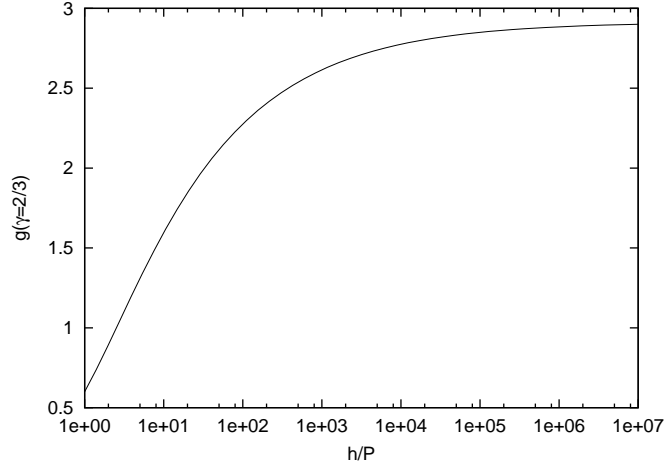}
\caption{The undersampling factor (\ref{eq.gambar}) as a function of the ratio $h/P$.
}
\label{fig.joInt}
\end{figure}

Collecting variances of optical path differences
from this model---obtained by striking the factor $\bar k^2$ of $\cal D_\varphi$---
we notice that the strength of the fluctuations remains smaller
than predicted by the factor $g(\gamma)$ in
(\ref{eq.gofgamm}), because this has been derived in
the limit $h/P\to \infty$ of infinite
layer height \cite{Goodman}. For finite $h/P$,
\begin{eqnarray}
{\cal D}_\varphi &=& 2\bar k^2 \int_0^h (h-\Delta z)\left[{\cal D}_n(\sqrt{(\Delta z)^2+P^2})-{\cal D}_n(\Delta z)\right]d(\Delta z)
\\
&=& 2\bar k^2 C_n^2 P^{\gamma+1} h \int_0^{h/P} \left(1-\frac{t}{h/P}\right)
\left[(t^2+1)^{\gamma/2}-t^\gamma \right]dt
,
\label{eq.gofP}
\end{eqnarray}
where the integral can be rephrased
as \cite[(3.243.1),(3.251.1)]{GR}
\begin{equation}
g(\gamma)= 2\int_0^{h/P} [1-t/(h/P)]\left[(1+t^2)^{\gamma/2}-t^\gamma\right]dt=
\frac{h}{P} \,_3F_2\left(\begin{array}{c}-\gamma/2,1/2,1 \\ 3/2,2\end{array}\mid -\left(\frac{h}{P}\right)^2\right) -\frac{(h/P)^{1+\gamma}}{(\gamma+1)(1+\gamma/2)}
,
\label{eq.gambar}
\end{equation}
with \cite{GottschalkJPA21,RainvilleBAMS51}
\begin{equation}
\,_3F_2\left(\begin{array}{c}-\gamma/2,1/2,1 \\ 3/2,2\end{array}\mid z\right)
=
2\,_2F_1\left(\begin{array}{c}-\gamma/2,1/2 \\ 3/2\end{array}\mid z\right)
-
\,_2F_1\left(\begin{array}{c}-\gamma/2,1 \\ 2\end{array}\mid z\right)
.
\end{equation}
This factor cannot reach $2.91$ in practise: the layer height $h$
multiplied by the air mass is limited to the order of 10 km representing the entire
atmosphere,
and $P$ is of the order of $2$ m for a single telescope up $100$ m for
an interferometer. Then $h/P$ is in the range 5000 to 100, and Figure \ref{fig.joInt}
demonstrates that this finite layer thickness places $g(\gamma)$ in the range 2.2 to 2.6\@.
Long baseline interferometers ``miss'' some power of ${\cal D}_\varphi$ through this
geometric sampling effect.

\section{Karhunen-Lo\`eve Modes in the Entrance Pupil}\label{sec.2d}
\subsection{Phase Structure Function in the Circular Pupil} \label{sec.p2d}
Removal of the spatial mean of the structure function (\ref{eq.Dphase})
defines associated covariances $C_\varphi$ for two-dimensional vectors
${\bf r}$ in the pupil plane and two-dimensional spatial wavenumbers $f$,
\begin{equation}
C_\varphi({\bf r})=\int C_\varphi(f)e^{-2\pi i{\bf f}\cdot {\bf r}} d^2f
=
2\pi \int_0^\infty C_\varphi(f)J_0(2\pi fr) fdf
,
\end{equation}
where
\begin{equation}
C_\varphi(f) = \bar k^2 h \Phi_n(f)
.
\label{eq.cphif}
\end{equation}
The statistically independent Karhunen-Lo\`eve (KL) modes $F_j$ are eigenvectors
of the integral operator \cite{FriedJOSA68}
\begin{equation}
\int_{|r'|<R} C_\varphi({\bf r}-{\bf r}') F_j({\bf r}') 
d^2r'
= {\cal B}^2 F_j({\bf r})
,
\end{equation}
where the integration is over the area limited by the pupil radius $R$.
Moving on to Fourier space, the convolution simplifies to
\begin{equation}
C_\varphi(f) 
F_j({\bf f}) = {\cal B}^2 F_j({\bf f}),
\end{equation}
and more explicitly with (\ref{eq.cphif}) to
\begin{equation}
\bar k^2 h c_n C_n^2
\frac{1}{(f^2+f_0^2)^{(\gamma+3)/2}}
F_j({\bf f}) = {\cal B}^2 F_j({\bf f})
.
\end{equation}
Introducing dimensionless variables ${\bf r}=R{\bf x}$ and $R{\bf f}=\bm{\sigma}$
helps to gather a set of common constants to unify the analysis,
\begin{equation}
 c_n
\frac{1}{(\sigma^2+\sigma_0^2)^{(\gamma+3)/2}}
K_p^{(q)}(\bm{\sigma}) = \frac{{\cal B}^2}{\bar k^2 h C_n^2 R^{\gamma+3}} K_p^{(q)}(\bm{\sigma})
.
\label{eq.Ksig}
\end{equation}
The notation assumes that azimuthal dependencies have been
split off,
\begin{equation}
K_p^{(q)}(\bm{\sigma}) = K_p^{(q)}(\sigma) M_q(\theta_\sigma),
\label{eq.KL2d}
\end{equation}
\begin{equation}
M_q(\theta)=\sqrt{\frac{\varepsilon_q}{2\pi}}\times\left\{\begin{array}{l}\cos(q\theta)\\
\sin(q\theta)\end{array}\right.
; \quad \varepsilon_q\equiv \left\{\begin{array}{rl}1, & q=0,\\
1,& |q|\ge 1,\end{array}\right.
\label{eq.Mtheta}
\end{equation}
such that the two discrete indices $p$ and $q$ can be used to enumerate
the
modes by
number of nodes in the radial direction and
periodicity along
the azimuth $\theta$.
If the $K_p^{(q)}(x)$ are expanded in the basis of Zernike polynomials $R_n^q$,
\begin{equation}
\int_0^1 x R_n^q(x)R_{n'}^q(x)dx = \frac{\delta_{n,n'}}{2(n+1)},
\end{equation}
\begin{equation}
K_p^{(q)}(x) = \sum_{n\equiv q \bmod 2} \tau_{n,p,q}\sqrt{2(n+1)} R_n^q(x),
\label{eq.Kpqx}
\end{equation}
Noll's Fourier representation of the $R_n^q$
induces a Fourier representation of the radial component
\cite{NollJOSA66,RoddierOE29},
\begin{equation}
K_p^{(q)}(\sigma)
= \sum_{n\equiv q \bmod 2} \tau_{n,p,q}\sqrt{2(n+1)} (-1)^{(n-q)/2} i^q
\frac{J_{n+1}(2\pi \sigma)}{\sigma}
= \sum_{n\equiv q \bmod 2} \tau_{n,p,q}\sqrt{2(n+1)} i^n
\frac{J_{n+1}(2\pi \sigma)}{\sigma}
\end{equation}
which becomes an eigenvalue equation for 
the vector of the expansion coefficients $\tau_{n,p,q}$ \cite{MatharArxiv0705b},
\begin{equation}
 c_n
(2\pi)^{\gamma+3}
\sum_{n\equiv q\bmod 2}
\tau_{n,p,q}\sqrt{2(n+1)}
i^{n-n'}\int_0^\infty
J_{n+1}(k) J_{n'+1}(k)
\frac{dk}{k(k^2+k_0^2)^{(3+\gamma)/2}}
= \frac{{\cal B}^2}{\bar k^2 h C_n^2 R^{\gamma+3}}
\tau_{n',p,q}\frac{1}{\sqrt{2(n'+1)}}
.
\end{equation}
At this stage some arbitrariness remains in scaling both sides of the equation,
depending on preferences of normalization to pupil areas,
squared diameters or radii. I shall divide through $g(\gamma)$
to gather $g(\gamma)$ in front of the $C_n^2$ of the right
hand side to match (\ref{eq.dKlim}),
and shall multiply by a factor
\begin{equation}
2c_\varphi
=
2\left[4\Gamma\left(\frac{3+\gamma}{1+\gamma}\right)\right]^{(1+\gamma)/2}
=
2\left[\frac{24}{5}\Gamma(6/5)\right]^{5/6}\approx 6.883877,\quad (\gamma=2/3)
\end{equation}
for compatibility with calculations that introduce a Fried radius as a reference.
Altogether,
\begin{equation}
-2c_\varphi \pi\frac{\Gamma(\frac{3}{2}+\frac{\gamma}{2})}{\Gamma(-\frac{1}{2}-\frac{\gamma}{2})}
\sum_{n\equiv q\bmod 2}
\tau_{n,p,q}\sqrt{(n+1)(n'+1)}
(-1)^{(n-n')/2}
I_{nn'}
= \lambda_{p,q}^2
\tau_{n',p,q}
\end{equation}
defining
\begin{equation}
\lambda_{p,q}^2 \equiv
\frac{2c_\varphi}{g(\gamma)} \frac{{\cal B}^2 }{\bar k^2 h C_n^2 (2R)^{\gamma+3}}
\label{eq.lam2d}
\end{equation}
and
\begin{equation}
I_{nn'}
\equiv
\int_0^\infty
J_{n+1}(k)
\frac{1}{k(k^2+k_0^2)^{(3+\gamma)/2}}
J_{n'+1}(k)
dk
.
\label{eq.Inndef}
\end{equation}
This integral is evaluated as proposed by Winker \cite{WinkerJOSAA8,TakatoJOSAA12},
\begin{eqnarray}
I_{nn'}
&=&
\frac{2}{\pi}\int_0^\infty dk
\int_0^{\pi/2} d\theta
J_{n+n'+2}(2k\cos\theta)\cos[(n-n')\theta]
\frac{1}{k(k^2+k_0^2)^{(3+\gamma)/2}}
\\
&=&
\frac{2}{\pi}\int_0^\infty dk
\int_0^{\pi/2} d\theta
\frac{1}{2\pi i}
\int_{-i\infty}^{i\infty} ds
\frac{\Gamma(-s)(k\cos\theta)^{n+n'+2+2s}}{\Gamma(n+n'+3+s)}
\cos[(n-n')\theta]
\frac{1}{k(k^2+k_0^2)^{(3+\gamma)/2}}
\end{eqnarray}
\begin{eqnarray}
&=&
\frac{2}{\pi}\int_0^\infty dk
\frac{1}{2\pi i}
\int_{-i\infty}^{i\infty} ds
\frac{\Gamma(-s)k^{n+n'+2+2s}}{\Gamma(n+n'+3+s)}
\frac{\pi\Gamma(n+n'+4+2s)}{2^{n+n'+3+2s}(n+n'+3+2s)\Gamma(n+2+s)\Gamma(n'+2+s)}
\frac{1}{k(k^2+k_0^2)^{(3+\gamma)/2}}
\nonumber
\\
&=&
k_0^{n+n'-1-\gamma}
\frac{\Gamma(\frac{1+\gamma-n-n'}{2})\Gamma(\frac{n+n'}{2}+1)}
{2^{n+n'+3}\Gamma(n+2)\Gamma(n'+2)\Gamma(\frac{3+\gamma}{2})}
\nonumber
\\
&&
\quad
\times
\,_3F_4(\frac{n+n'}{2}+1,\frac{n+n'}{2}+2,\frac{n+n'+3}{2} ;
n+2,n'+2,n+n'+3,\frac{1+n+n'-\gamma}{2}
;k_0^2) 
\nonumber
\\
&&
+
\frac{\Gamma(\frac{n+n'-1-\gamma}{2})\Gamma(4+\gamma)}
{2^{\gamma+4}\Gamma(\frac{7+\gamma+n+n'}{2})
\Gamma(\frac{5+\gamma+n-n'}{2})\Gamma(\frac{5+\gamma+n'-n}{2})}
\nonumber
\\
&&
\quad
\times
\,_3F_4(\frac{5+\gamma}{2},\frac{3+\gamma}{2},2+\frac{\gamma}{2} ;
\frac{5+\gamma+n-n'}{2},\frac{5+\gamma+n'-n}{2},
\frac{7+\gamma+n+n'}{2},\frac{3+\gamma-n-n'}{2}
;k_0^2) 
.
\label{eq.Inn}
\end{eqnarray}

\begin{figure}[hbt]
\includegraphics[scale=0.6]{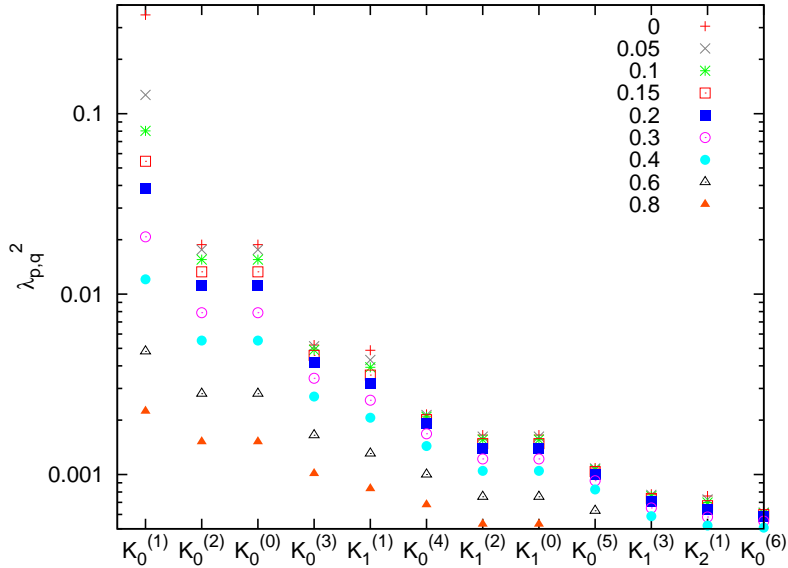}
\caption{
The eigenvalues $\lambda_{p,q}^2$ (\ref{eq.lam2d}) of the KL modes of the
phase variance as a function of the outer scale wavenumber $0\le \sigma_0 \le 0.8$.
}
\label{fig.showEvl}
\end{figure}

\begin{figure}[hbt]
\includegraphics[scale=0.4]{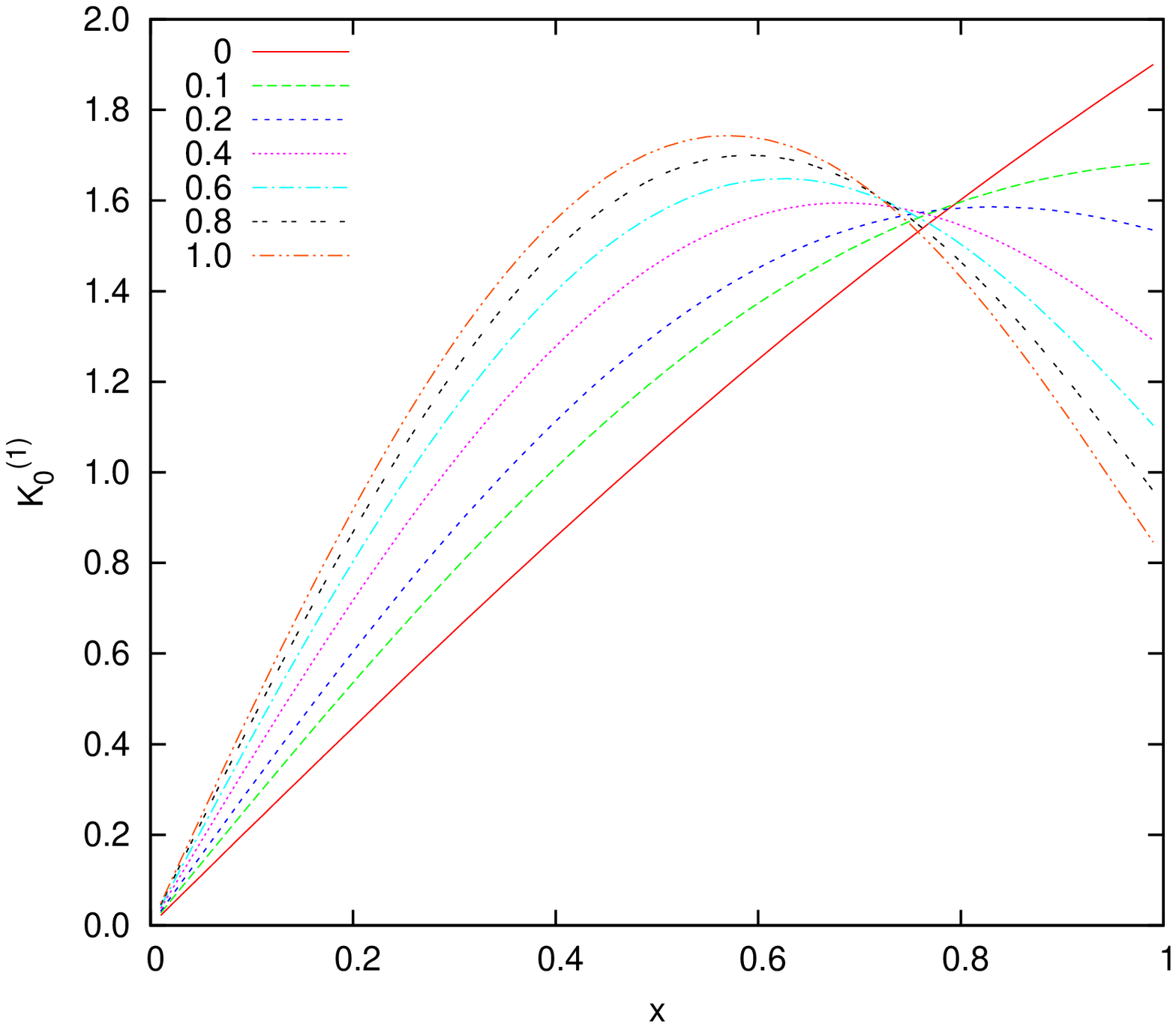}
\includegraphics[scale=0.4]{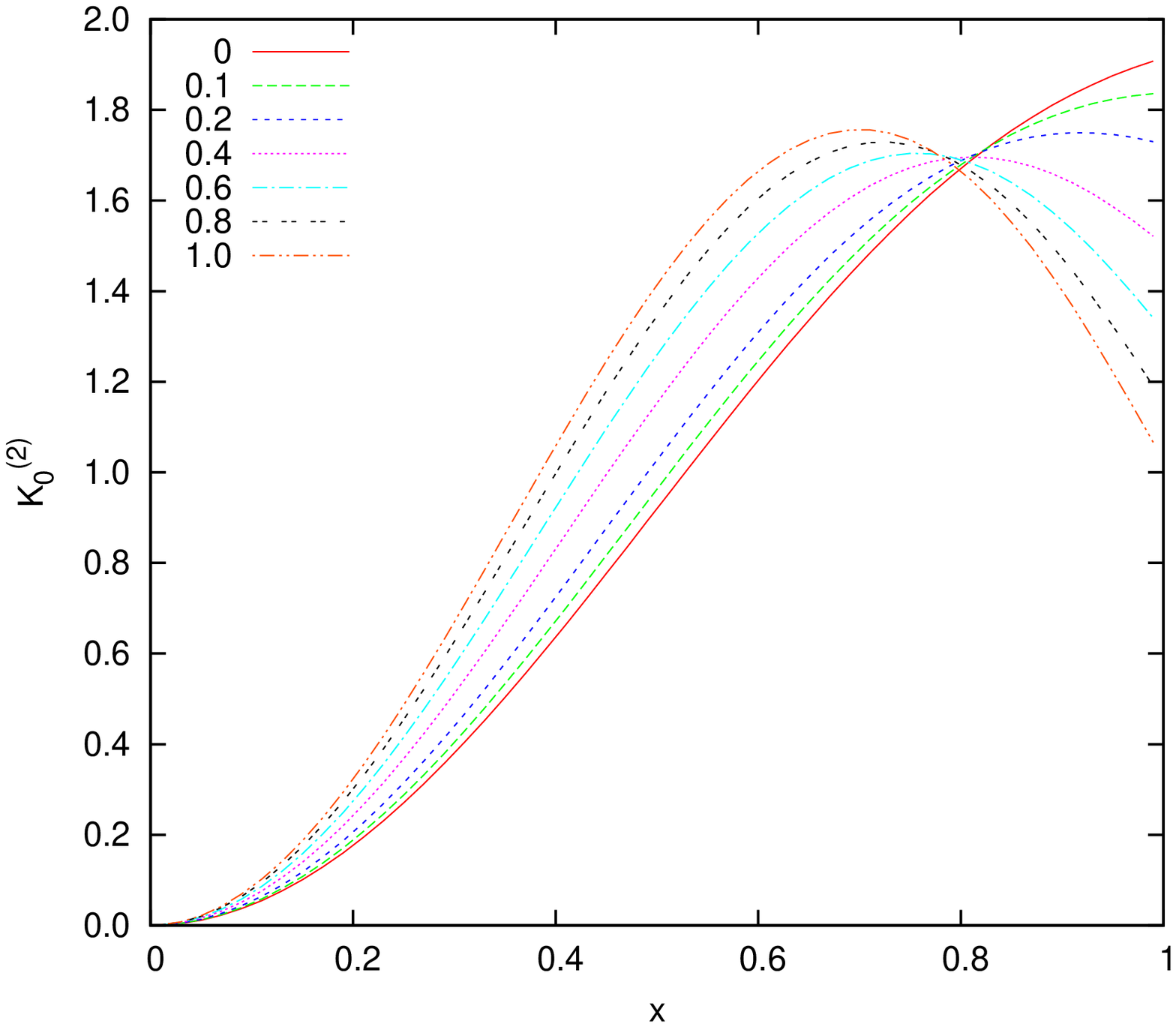}
\includegraphics[scale=0.4]{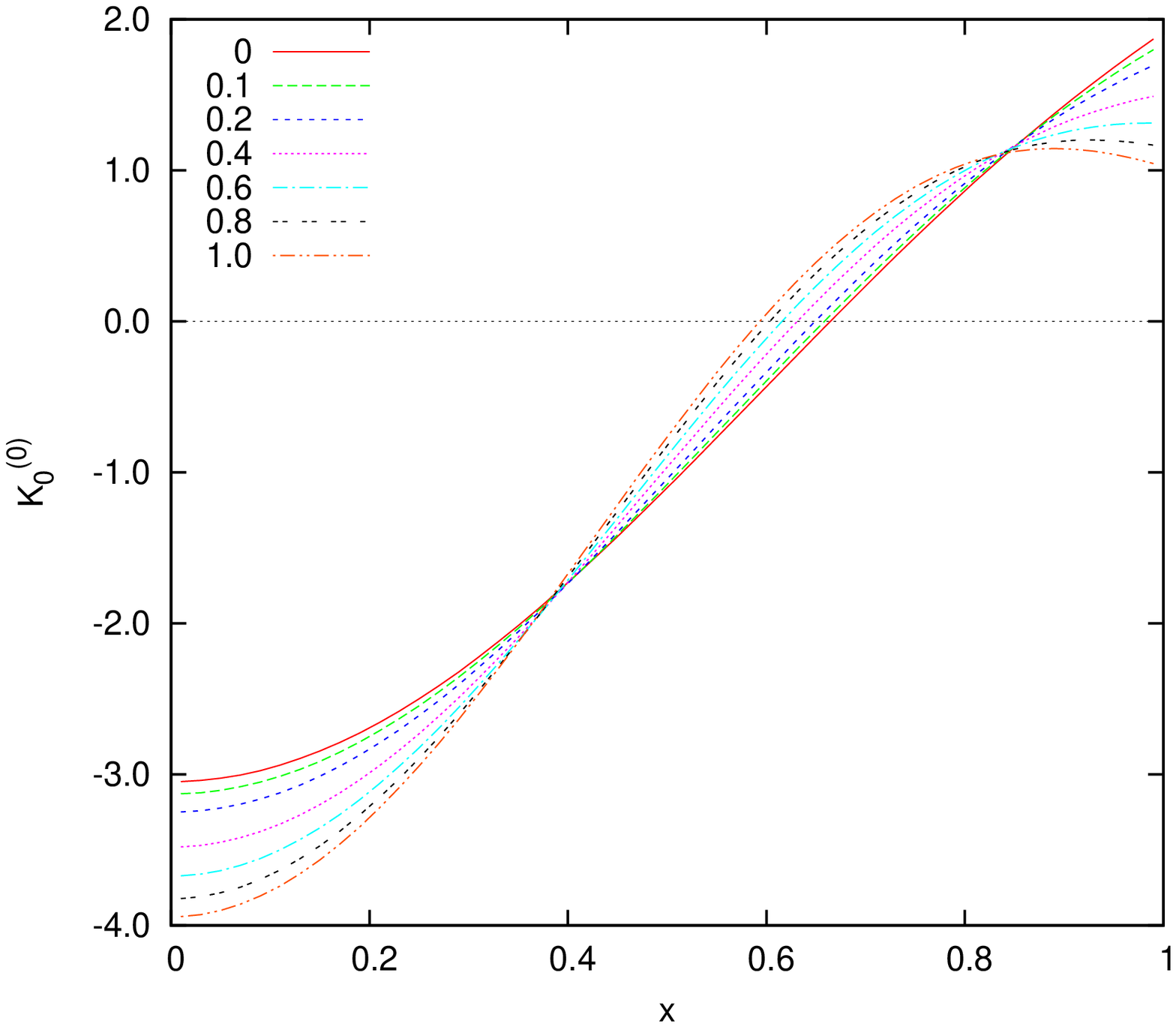}
\includegraphics[scale=0.4]{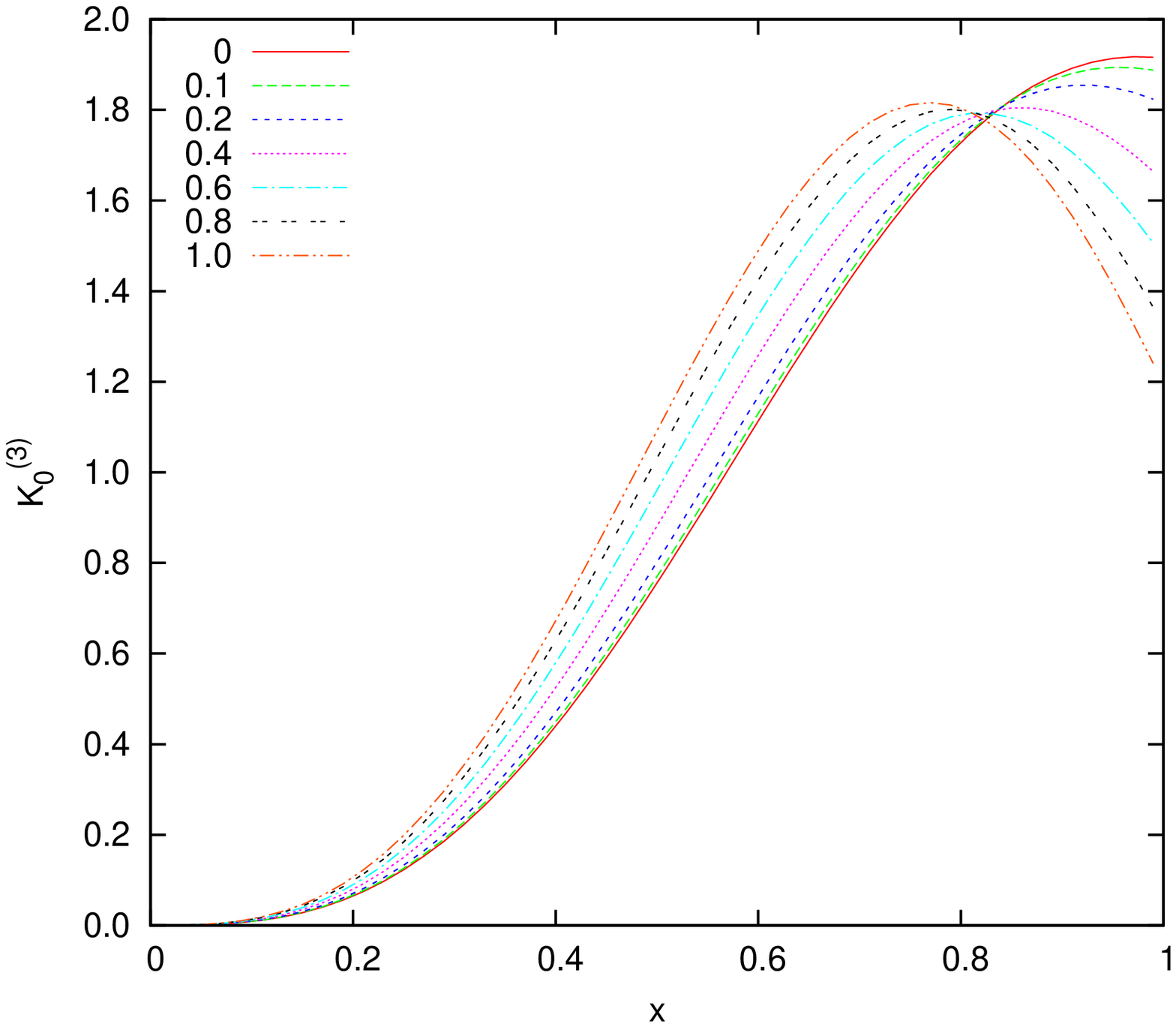}
\includegraphics[scale=0.4]{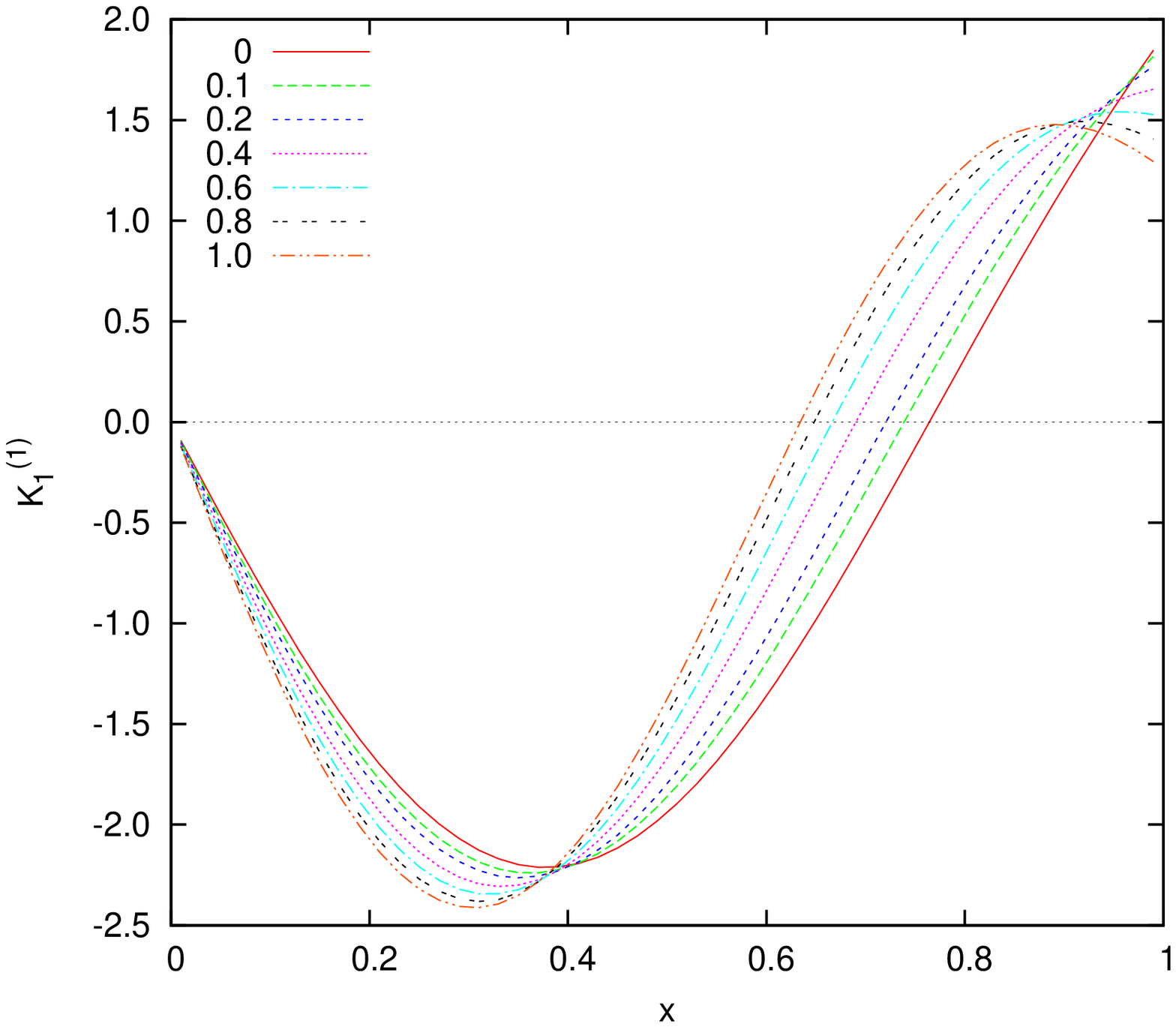}
\includegraphics[scale=0.4]{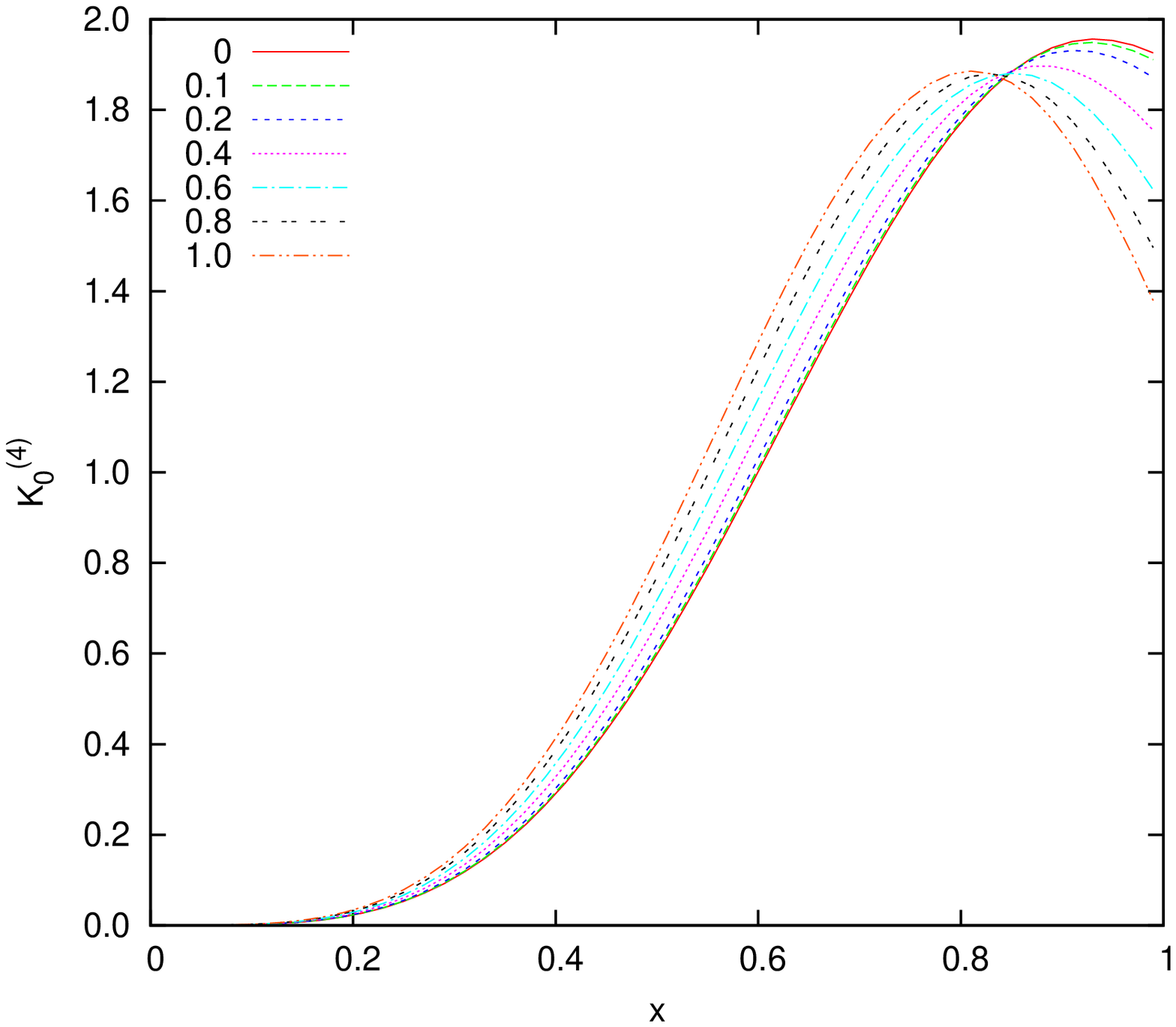}
\caption{Influence of the inverse outer scale $0\le \sigma_0\le 1$
on some large-scale KL modes (\ref{eq.Kpqx}) of the
phase covariance in the circular input pupil.
}
\label{fig.KLroddi}
\end{figure}

Prototypical results of the calculation are listed in App.\ \ref{app.kl2d}.
The reduction of the eigenvalues as the outer scale shrinks is visualized
in Fig.\ \ref{fig.showEvl}, and the inward motion
of the normalized radial functions $K_p^{(q)}(x)$ is shown in Fig.\ \ref{fig.KLroddi}.
The modes at the largest scales are effected most,
as known from earlier analyses \cite{WinkerJOSAA8}.

\subsection{Estimation of Outer Scales} \label{sec.meth}

The customary reduction of AO data to outer scales reconstructs
the phase as a function of the location in the pupil plane,
collects the expansion coefficients of the Zernike expansion,
and compares the relative strengths of the coefficients with
the reduction predicted as a function of the outer scale
\cite{FuscoSPIE5490,FuscoJoptA6,WinkerJOSAA8}.

Which new aspects arise if a chart of the KL modes as in Figure \ref{fig.showEvl}
is used instead?

Obtaining the expansion coefficients still is a matter of fitting a phase
function to a set of modes. One may start with a Zernike basis \cite{MalacaraOE29}
and convert its coefficients to the KL basis with the aid of Appendix \ref{app.kl2d}.
The appendix provides the matrix that converts from the Zernike to the KL
basis, but no additional computation is demanded because
the inverse matrix is the transpose matrix. At this step,
the tip-tilt component is an essential ingredient to the process---this is
a requirement that was not known to the approach based on the
Zernike covariances, and probably disables processing if tip-tilt
stages are autonomous and data not available.
If a modular approach to information processing of various layers of
phase corrections in telescope operation---from course with fringe tracking
over medium of tip-tilt stages to fine of adaptive optics---results
in a lean system optimized to dispose intermediate data streams irrelevant
to the observing astronomer, post-processing of the engineering data concerning
components like the atmosphere is put on a diet.

Another modest technical disadvantage is that 
fitting process becomes iterative in nature because the entries
of Appendix \ref{app.kl2d} depend on the outer scale, which is the unknown.
This is only a matter of producing this static table for a range of 
outer scales of interest and therefore no real concern.

The conceptual difference is that the data reduction is based
on the statistical independence of the KL modes. The outer
scale is
selected
that generates measured variances of the modal coefficients
proportional to their $\lambda_{p,q}^2$. (The overall factor $C_n^2$ drops out
of the calculation.) The technique  does not derive the outer scale
from any individual snapshot, but it starts from a hypothetical outer scale,
reduces the series of exposures to variances of modal coefficients, and readjusts
the value of the outer scale until these variances match that common statistics.
In addition, standard statistical methodologies can be applied to derive
confidence intervals of the result.

\subsection{Central Obscuration} \label{sec.obsc}
A central obscuration from a secondary mirror removes data up to
a radius $\epsilon$ [measured in units of the pupil radius,  $0\le \epsilon\le 1$, unrelated
to the $\varepsilon$ of (\ref{eq.Mtheta})] from the areal
statistics.
The radial component of (\ref{eq.KL2d}) is represented by the Fourier pair
\begin{equation}
K_p^{(q)}(\sigma)=2\pi i^q\int_\epsilon^1 xdx K_p^{(q)}(x)J_q(2\pi\sigma x)
,\quad
K_p^{(q)}(x)=2\pi (-i)^q\int_0^\infty  \sigma d\sigma K_p^{(q)}(\sigma)J_q(2\pi\sigma x)
,
\label{eq.kpqeps}
\end{equation}
which recovers the results of the previous section if $\epsilon=0$. Insertion of $K_p^{(q)}(\sigma)$
in (\ref{eq.Ksig}) yields the eigenvalue equation in mixed coordinates,
\begin{equation}
2\pi (-i)^q c_n
\int_0^\infty \frac{1}{(\sigma^2+\sigma_0^2)^{(\gamma+3)/2}}
\sigma K_p^{(q)}(\sigma) J_q(2\pi\sigma x)d \sigma=
\frac{{\cal B}^2}{\bar k^2 h C_n^2 R^{\gamma+3}} K_p^{(q)}(x)
.
\end{equation}
Insertion of (\ref{eq.kpqeps}) and scaling the wave number with $k\equiv 2\pi\sigma$
means
\begin{equation}
(2\pi)^{\gamma+3} c_n
\int_0^\infty dk \frac{1}{(k^2+k_0^2)^{(\gamma+3)/2}}
\int_\epsilon^1 x' dx' K_p^{(q)}(x') k J_q(k x') J_q(k x)=
\frac{{\cal B}^2}{\bar k^2 h C_n^2 R^{\gamma+3}} K_p^{(q)}(x)
.
\label{eq.evHybrid}
\end{equation}
The integral over $k$ is managed by isolating variables
\cite{BaileyPCPS25}\cite[(\S 7.15)]{Erdelyi}
\begin{eqnarray}
\frac{k}{2} J_q(kx) J_q(kx')
&=& (xx')^q\sum_{n\ge 0} (-1)^n(2q+2n+1)J_{2q+2n+1}(k)
{2q+n \choose q} {q+n \choose q}
\nonumber
\\
&& \times
\,_2F_1\left(\begin{array}{c}-n,2q+n+1\\ q+1\end{array}\mid \sin^2\phi\right)
\,_2F_1\left(\begin{array}{c}-n,2q+n+1\\ q+1\end{array}\mid \sin^2\Phi\right)
.
\label{eq.Bail}
\end{eqnarray}
Here, the two auxiliary variables $\phi$ and $\Phi$ are defined by
\begin{equation}
\cos\phi \cos\Phi = x;\quad 
\sin\phi \sin\Phi = x',
\end{equation}
that is
\begin{eqnarray}
\sin^2\phi &=& 
\frac{1- (x^2-x'^2) +\sqrt{1-(x+x')^2}\sqrt{1-(x-x')^2} }{2},\\
\sin^2\Phi &=& 
\frac{1- (x^2-x'^2) -\sqrt{1-(x+x')^2}\sqrt{1-(x-x')^2} }{2}.
\label{eq.Bailvars}
\end{eqnarray}
The two Hypergeometric Functions in (\ref{eq.Bail}) are Jacobi Polynomials \cite[(22.5.42)]{AS}.
Each term is then integrated with \cite[6.565.8]{GR}
\begin{eqnarray}
\int_0^\infty \frac{ J_{2q+2n+1}(k)}{(k^2+k_0^2)^{(\gamma+3)/2}}dk
&
=
&
\frac{k_0^{2q+2n-\gamma-1}\Gamma(1+q+n)\Gamma(\frac{\gamma+1}{2}-q-n)}{4^{q+n+1}
\Gamma(\frac{\gamma+3}{2})\Gamma(2q+2n+2)}
\,_1F_2\left(\begin{array}{c}1+q+n \\ q+n+\frac{1-\gamma}{2} , 2n+2q+2\end{array}\mid \frac{k_0^2}{4}\right)
\nonumber
\\
&&
+
\frac{\Gamma(n+q-\frac{\gamma+1}{2})}{2^{\gamma+3}\Gamma(\frac{\gamma+5}{2}+q+n)}
\,_1F_2\left(\begin{array}{c}\frac{\gamma+3}{2} \\ \frac{\gamma+5}{2}+q+n, \frac{\gamma+3}{2} -q-n\end{array}\mid \frac{k_0^2}{4}\right)
.
\end{eqnarray}
To avoid that
$x+x'>1$
in (\ref{eq.Bailvars}),
another variable transform $t=2k$ is added.
The following results
have been obtained with a finite element decomposition of the interval $[\epsilon,1]$,
and introducing the function
\begin{equation}
\tilde K_p^{(q)}(x)\equiv \sqrt{x/2}K_p^{(q)}(x),
\end{equation}
such that the matrix on the left hand side, which couples each $x$ to all the
$x'$, becomes symmetric:
\begin{equation}
(2\pi)^{\gamma+3} \frac{c_n c_\varphi}{g(\gamma)}
\int_0^\infty dt \frac{1}{(t^2+4k_0^2)^{(\gamma+3)/2}}
\int_\epsilon^1 dx' \sqrt{xx'} \tilde K_p^{(q)}(x') \frac{t}{2} J_q(t x'/2) J_q(t x/2)=
\frac{2c_\varphi}{g(\gamma)}\frac{{\cal B}^2}{\bar k^2 h C_n^2 (2R)^{\gamma+3}} \tilde K_p^{(q)}(x)
.
\end{equation}
The piston mode has been reduced
as outlined in Appendix \ref{app.pist}.

\begin{figure}[hbt]
\includegraphics[scale=0.4]{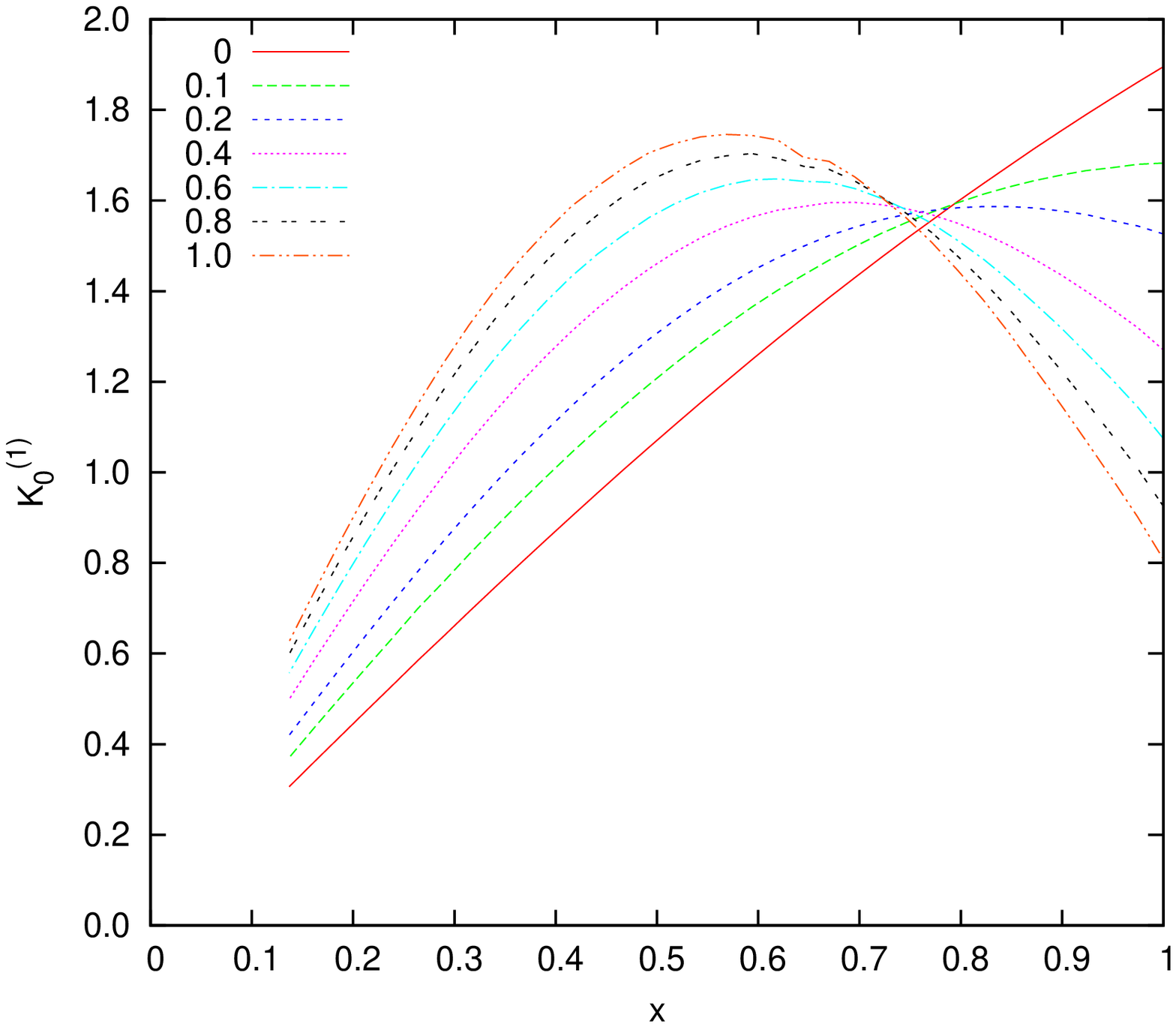}
\includegraphics[scale=0.4]{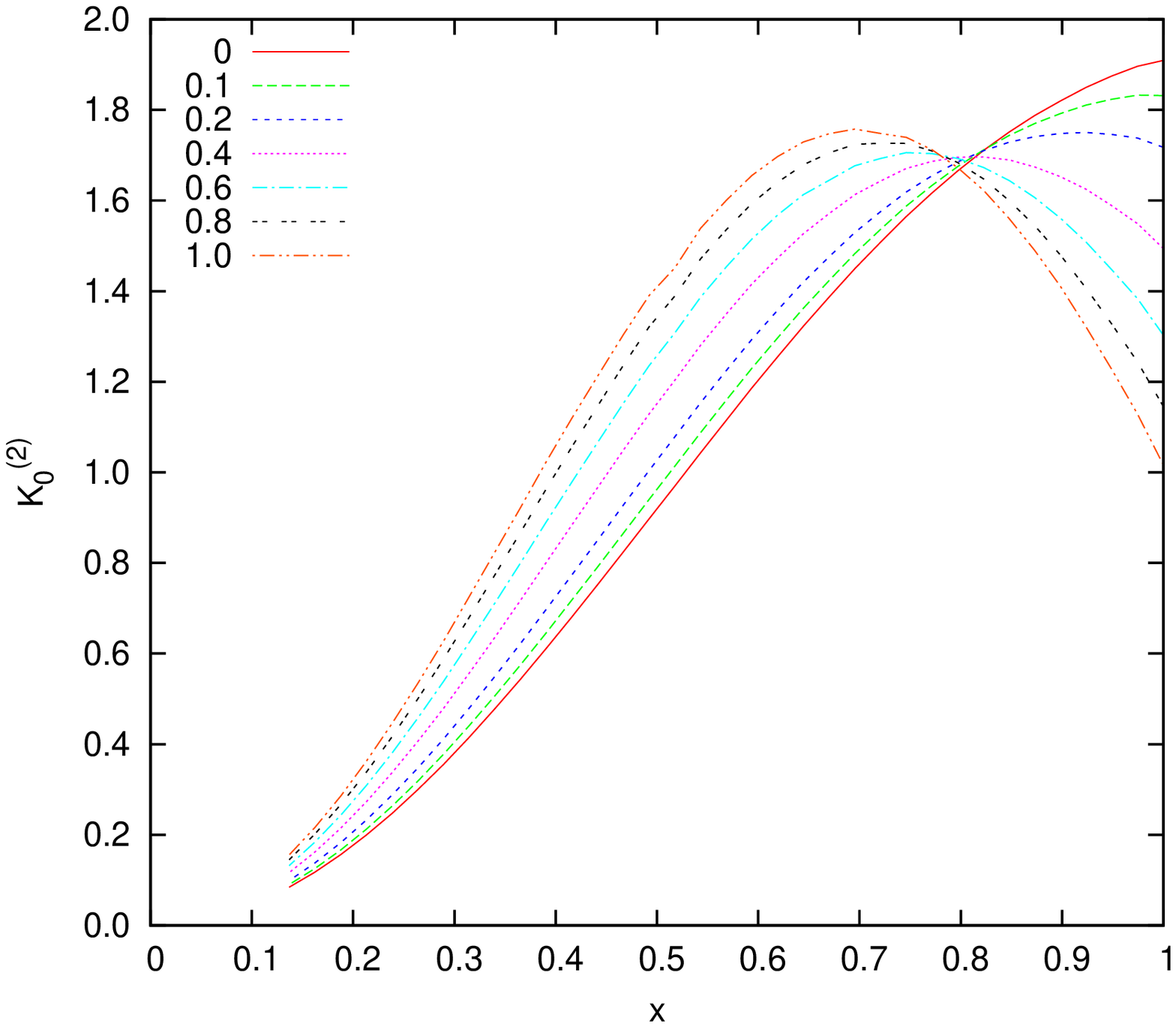}
\includegraphics[scale=0.4]{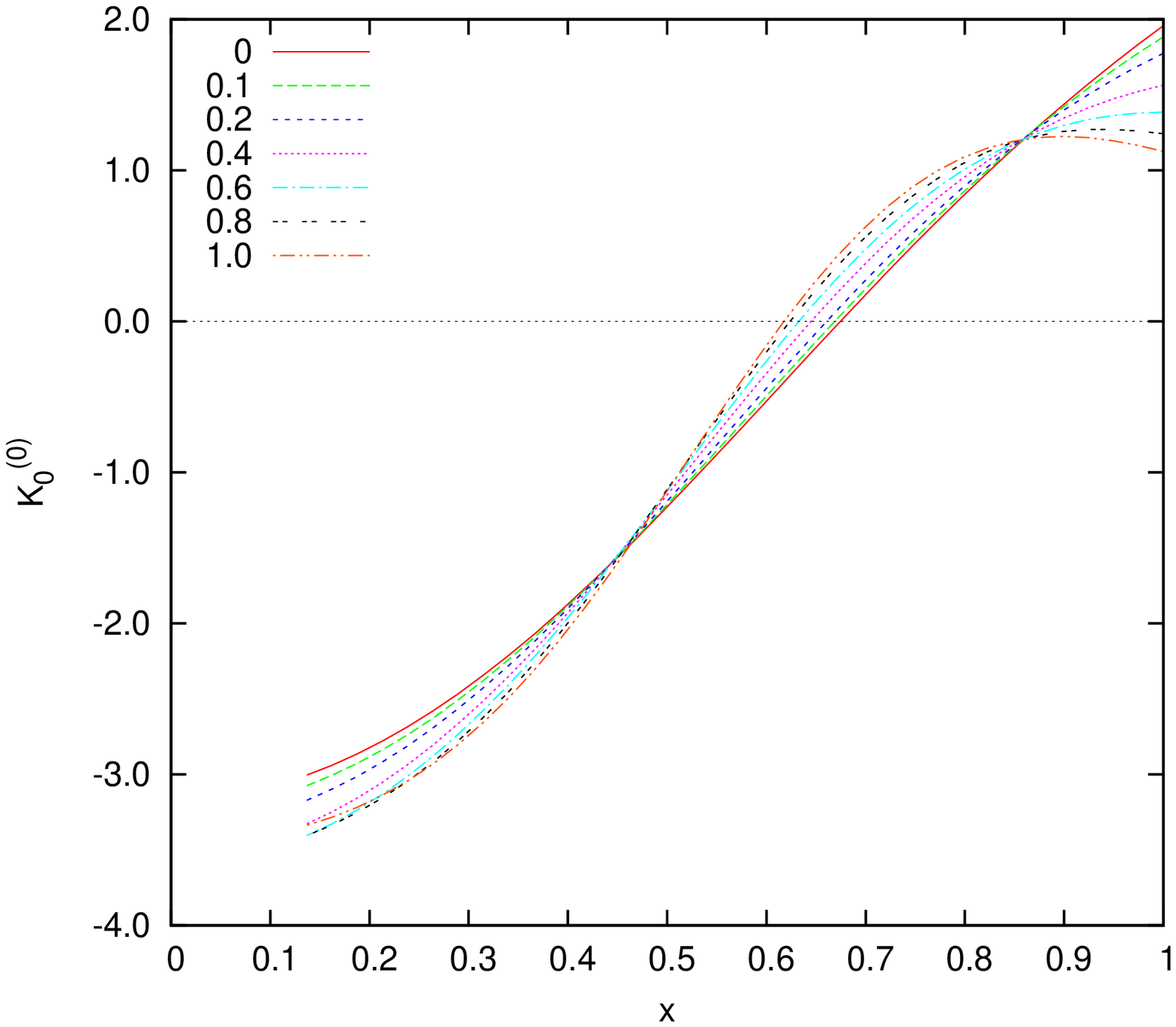}
\includegraphics[scale=0.4]{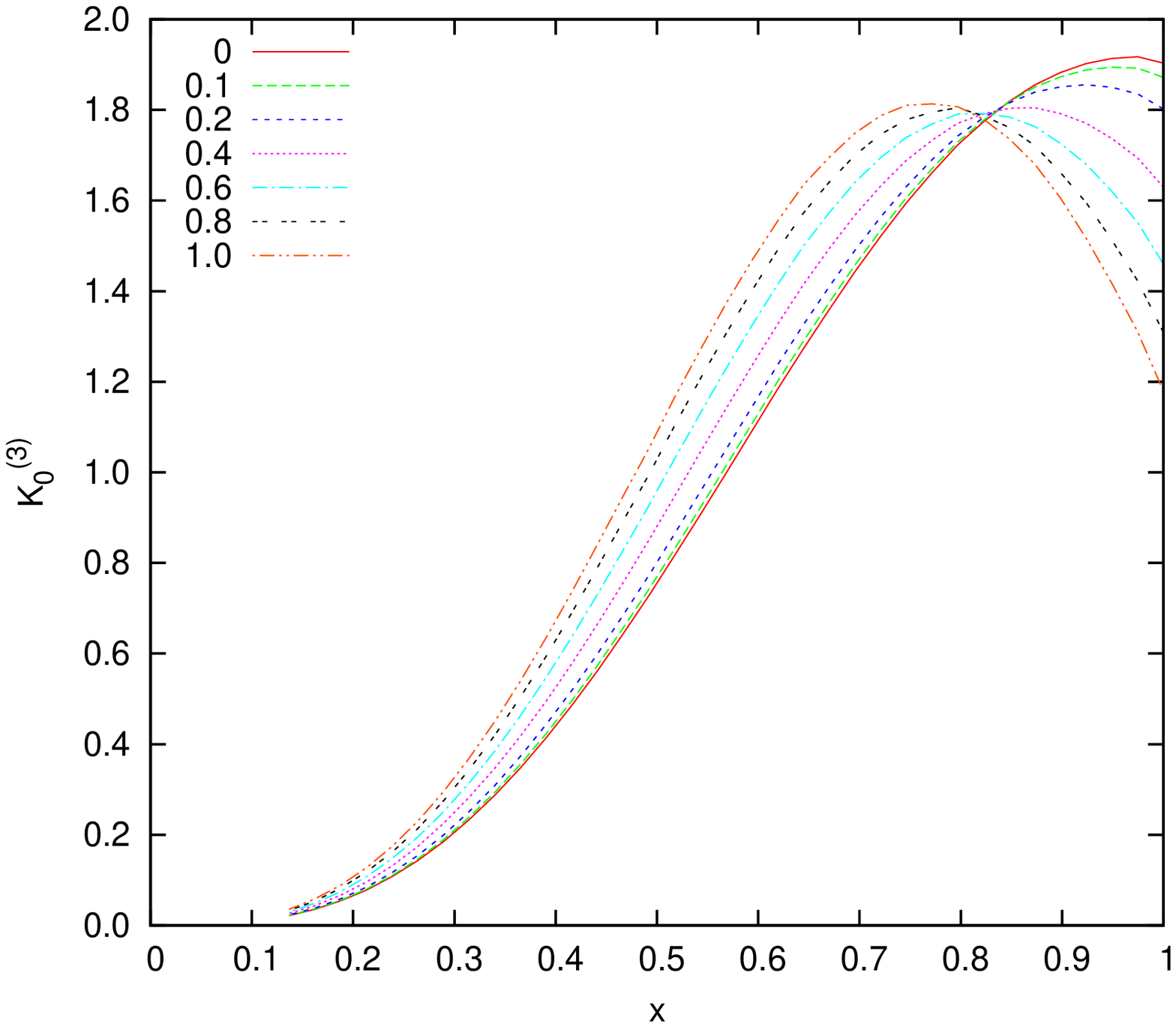}
\includegraphics[scale=0.4]{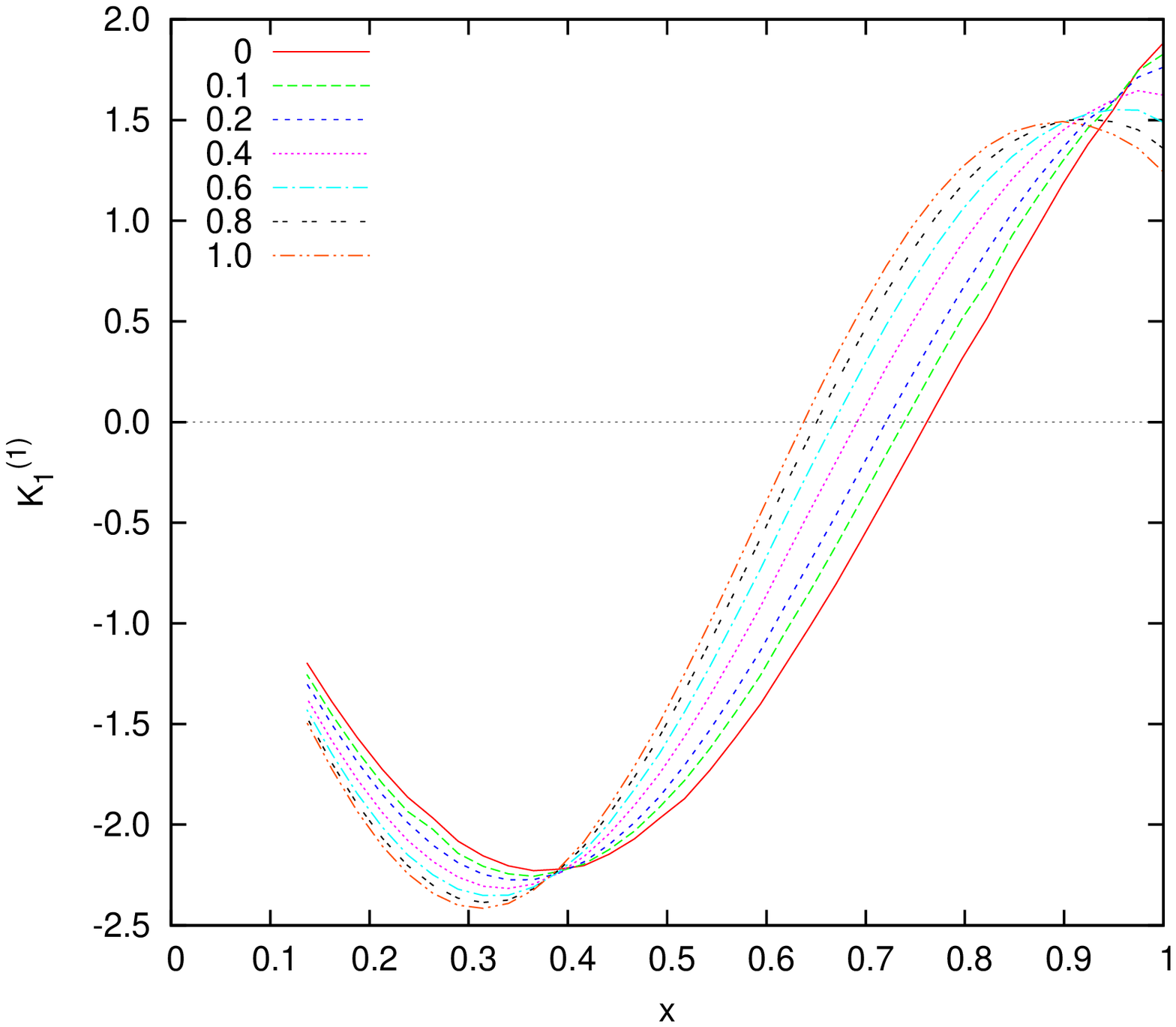}
\includegraphics[scale=0.4]{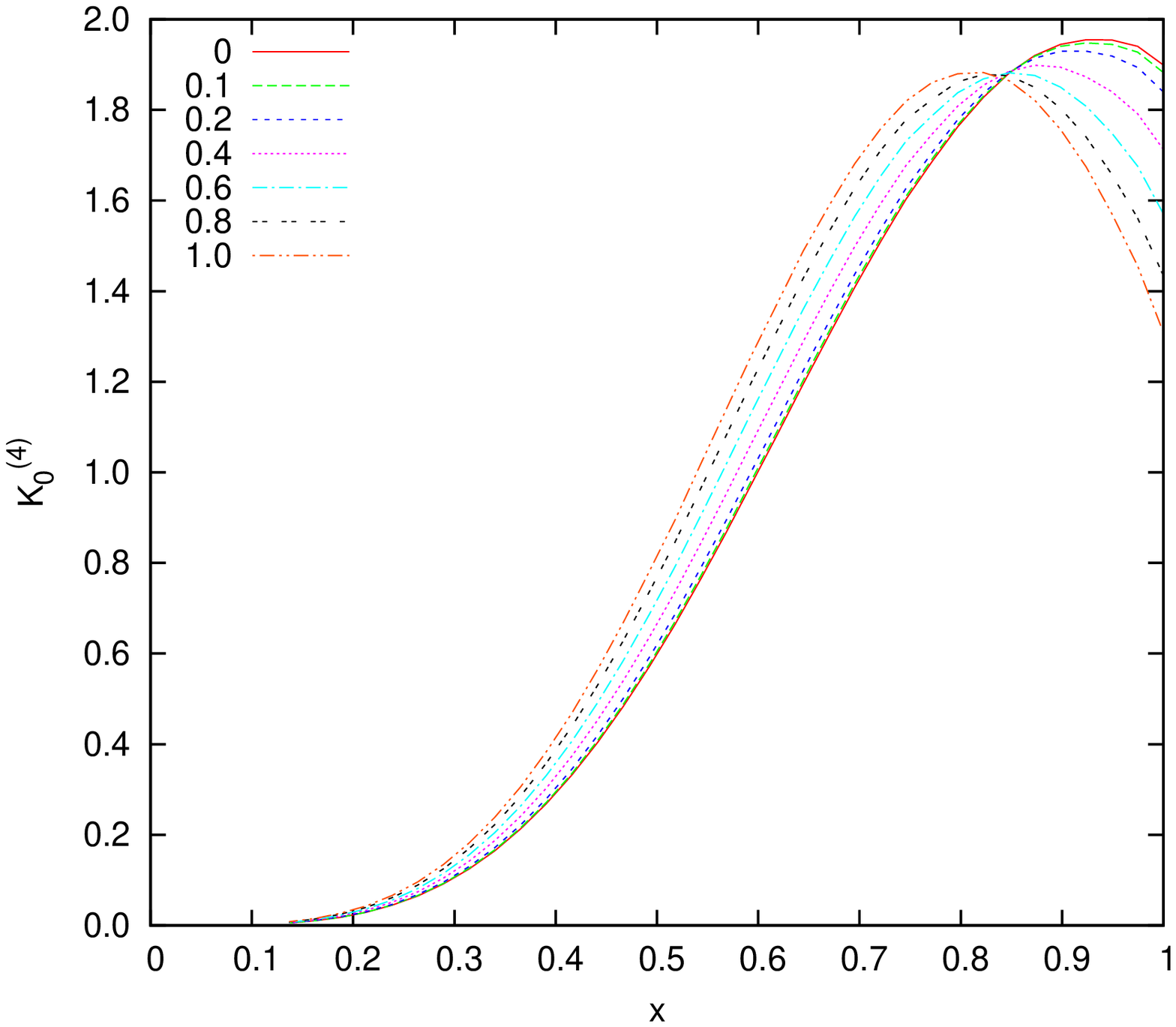}
\caption{
Large-scale KL modes $K_p^{(q)}(x)$ with a central obscuration of radius $\epsilon=0.1369$
for a set of inverse outer scales from 0 to 1.0.
}
\label{fig.obsc02}
\end{figure}

If the radius is selected equivalent
to the M2/M1 ratio of the telescopes of the Very Large Telescope, Figure \ref{fig.obsc02}
results. Besides the absence of values in the range $x\le \epsilon$,
these shapes are hardly different from the full aperture modes of Figure \ref{fig.KLroddi}.

\begin{figure}[hbt]
\includegraphics[scale=0.5]{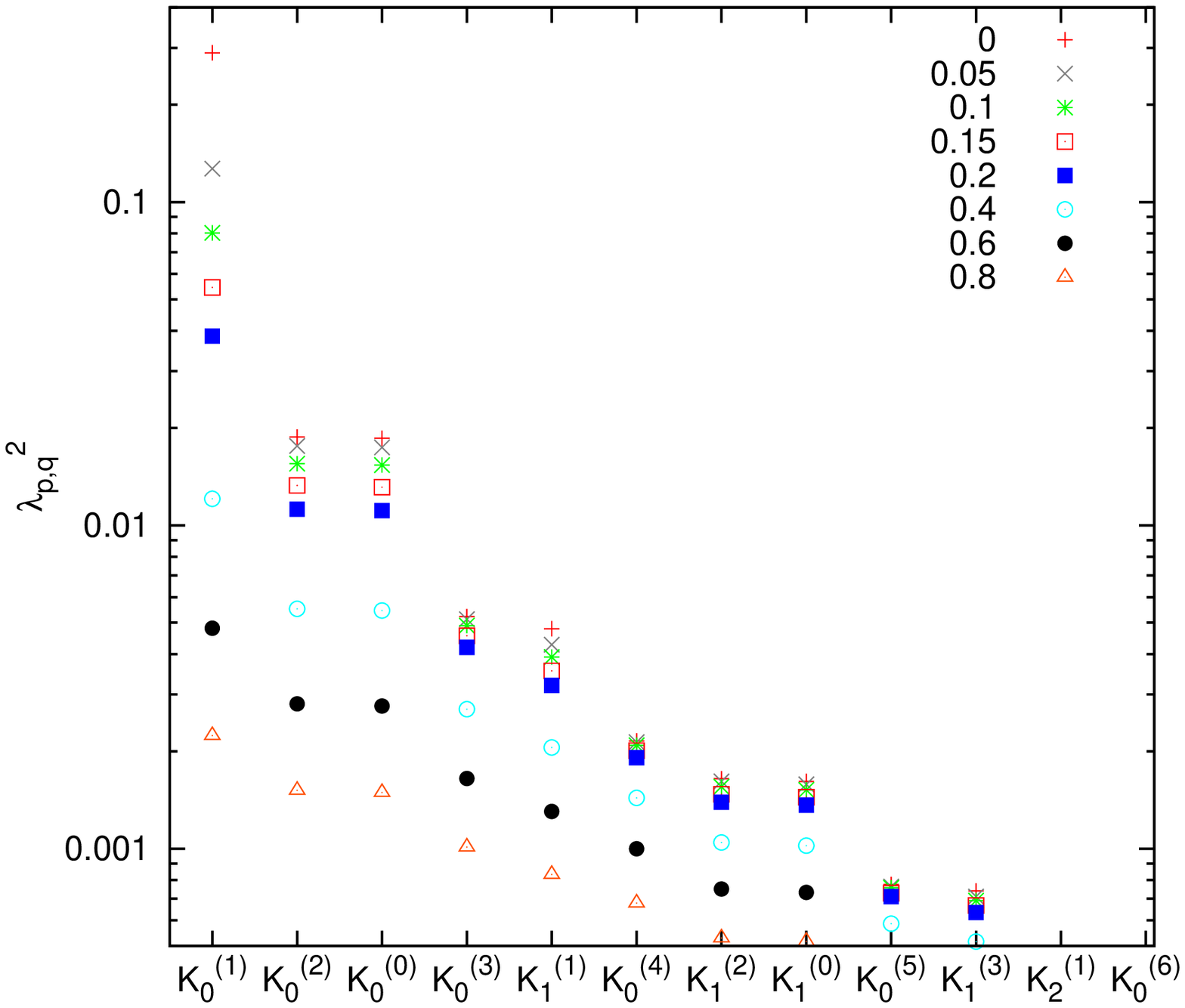}
\includegraphics[scale=0.5]{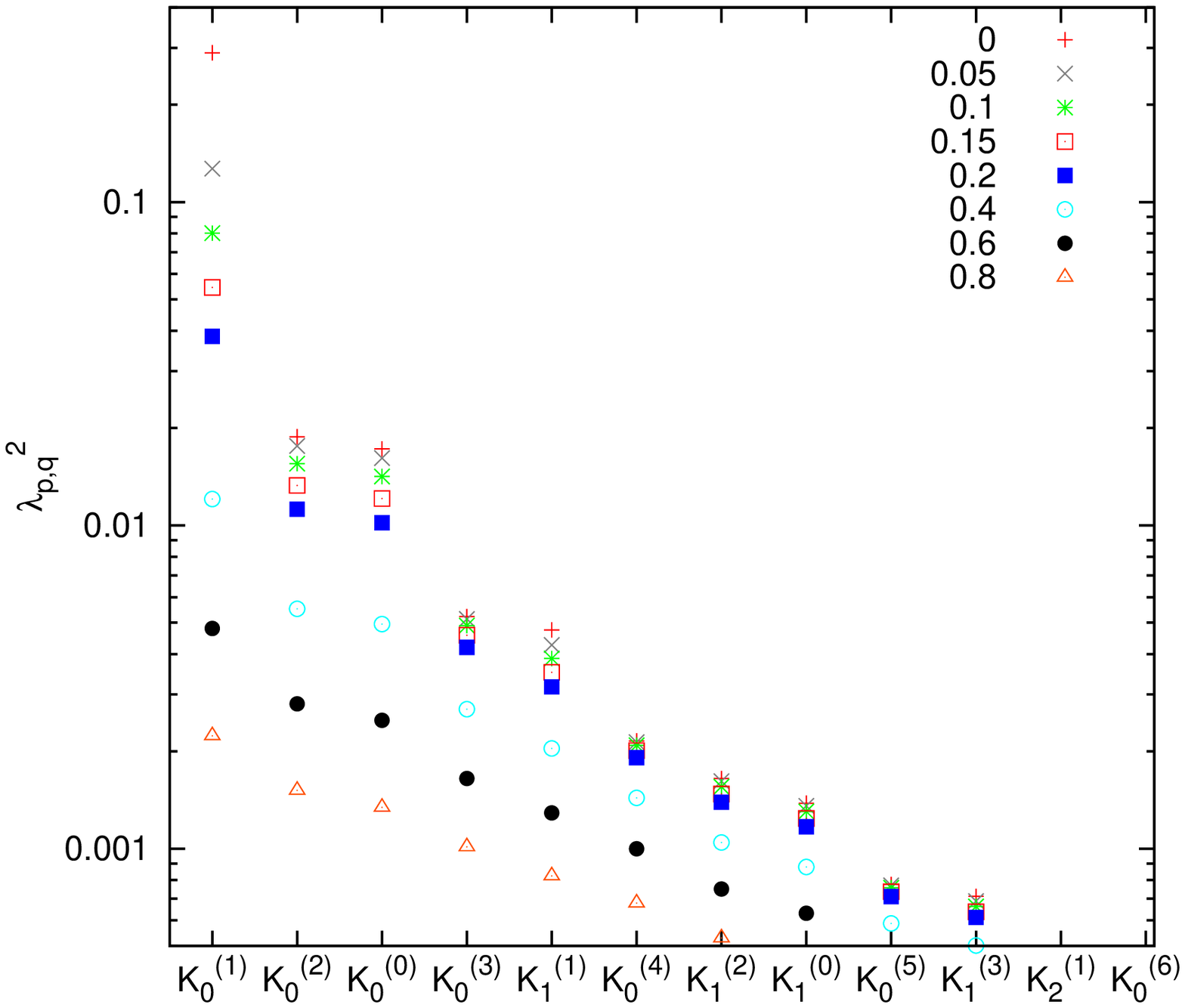}
\caption{
Modal eigenvalues for $\epsilon=0.0462$ (left, roughly representative of the Keck center
tile size) and $\epsilon=0.1369$ (right) with different symbols for inverse outer scales $\sigma_0$
from 0 to 0.8.
}
\label{fig.obl}
\end{figure}

The variances $\lambda^2$ in the individual modes 
are shown in Figure \ref{fig.obl}. [They are still defined by (\ref{eq.lam2d}),
so comparison of values for different $\epsilon$ is problematic.
One might argue that the reduction of the sampling area and the related
decrease in value would warrant a scaling
that should depend in one form or the other on $\epsilon$. Ultimately, this kind of
manipulation is futile, because the $\lambda^2$ do not depend only on
the pupil area but also on its shape.]
A noticeable effect is the break-up of the degeneracy of the
values for $K_p^{(2)}$ and $K_p^{(0)}$ as $\epsilon$ increases. It is plausible
that modes that do not have a node in the middle of the pupil
are more sensitive to
blanking of the center part than modes that approach zero as $x\to 0$.

\section{Summary} 
The text demonstrates computation of the statistically independent modes
of phase screens in the von K\'arm\'an class of outer scales sampled
over spherical receiver pupils, optionally taking into account masking
of a circular region in the center.
The main value in this work resides in techniques to compute
integrals in the eigenvalue problem.

These modes can be employed to compute outer scales from measured phase
functions. The goodness-of-fit of
the von K\'arm\'an modal spectrum
can be addressed.
The number of degrees of freedom depends on how
many of the modes remain free from other genuine sources of noise besides
atmospheric turbulence.

\appendix
\section{Karhunen-Lo\`eve Modes of the Phase Covariance Inside the Circular Pupil} \label{app.kl2d}
Explicit expansions of the first (large-scale) modes (\ref{eq.KL2d})
follow for dimensionless inverse outer scales from $0.05$ to $1$.
The case of infinite outer scale, $\sigma_0=0$, has been
published before
\cite{MatharArxiv0705b} and is not duplicated here.
Each line shows the eigenvalue (\ref{eq.lam2d}), a colon, the azimuthal parameter $|q|$,
another colon, and a sum over Zernike functions using Noll's indexing \cite{NollJOSA66}.
The $\sin(q\theta)$ and $\cos(q\theta)$ are basically degenerate if $q\neq 0$:
two modes share the same radial $K_p^{(q)}(x)$ and $\lambda_{p,q}^2$ and differ
only in $M_q(\theta)$.
A notation with two comma-separated indices for $Z$ condenses them
into a single line, and choosing either the left or the right number in the index pairs selects
a unique $K_p^{(q)}({\bf x})$.

The squared coefficients sum up to $1/\pi$ to cancel Noll's normalization of the
Zernike basis and to end up with a $K_p^{(q)}$ normalized to unity.

$\sigma_0=0.05$:
\small \begin{verbatim}
0.1270639 :  1:  +0.5621830*Z(2,3) -0.0474462*Z(8,7) +0.0030025*Z(16,17) -0.0000182*Z(30,29) 
+0.0000032*Z(46,47)

0.0176247 :  2:  +0.5540843*Z(6,5) -0.1056709*Z(12,13) +0.0115730*Z(24,23) -0.0004187*Z(38,39) 
+0.0000243*Z(58,57) +0.0000014*Z(80,81)

0.0176247 :  0:  +0.5540843*Z(1) -0.1056709*Z(4) +0.0115730*Z(11) -0.0004187*Z(22) +0.0000243*Z(37) 
+0.0000014*Z(56)

0.0051337 :  3:  +0.5411529*Z(10,9) -0.1576676*Z(18,19) +0.0245247*Z(32,31) -0.0016922*Z(48,49) 
+0.0001077*Z(70,69)

0.0043151 :  1:  +0.0457305*Z(2,3) +0.5301453*Z(8,7) -0.1848336*Z(16,17) +0.0315406*Z(30,29) 
-0.0025078*Z(46,47) +0.0001649*Z(68,67) -0.0000020*Z(92,93) +0.0000011*Z(122,121)

0.0021374 :  4:  +0.5271594*Z(14,15) -0.1971752*Z(26,25) +0.0389824*Z(40,41) -0.0038784*Z(60,59) 
+0.0003054*Z(82,83) -0.0000094*Z(110,109) +0.0000018*Z(140,141)

0.0016213 :  2:  +0.0990281*Z(6,5) +0.4914390*Z(12,13) -0.2517952*Z(24,23) +0.0594666*Z(38,39) 
-0.0073175*Z(58,57) +0.0006477*Z(80,81) -0.0000312*Z(108,107) +0.0000032*Z(138,139)

0.0016213 :  0:  +0.0990281*Z(1) +0.4914390*Z(4) -0.2517952*Z(11) +0.0594666*Z(22) -0.0073175*Z(37) 
+0.0006477*Z(56) -0.0000312*Z(79) +0.0000032*Z(106)

0.0010797 :  5:  +0.5135000*Z(20,21) -0.2273859*Z(34,33) +0.0536290*Z(50,51) -0.0068392*Z(72,71) 
+0.0006548*Z(96,97) -0.0000348*Z(126,125) +0.0000037*Z(158,159)

0.0007713 :  3:  +0.1435201*Z(10,9) +0.4445464*Z(18,19) -0.3025397*Z(32,31) +0.0912672*Z(48,49) 
-0.0150800*Z(70,69) +0.0017271*Z(94,95) -0.0001286*Z(124,123) +0.0000102*Z(156,157)

\end{verbatim}\normalsize

$\sigma_0=0.1$:
\small \begin{verbatim}
0.0802920 :  1:  +0.5603194*Z(2,3) -0.0658073*Z(8,7) +0.0046317*Z(16,17) -0.0000506*Z(30,29) 
+0.0000052*Z(46,47)

0.0155178 :  2:  +0.5522704*Z(6,5) -0.1146220*Z(12,13) +0.0129920*Z(24,23) -0.0005054*Z(38,39) 
+0.0000290*Z(58,57) +0.0000014*Z(80,81)

0.0155178 :  0:  +0.5522704*Z(1) -0.1146220*Z(4) +0.0129920*Z(11) -0.0005054*Z(22) +0.0000290*Z(37) 
+0.0000014*Z(56)

0.0049007 :  3:  +0.5400346*Z(10,9) -0.1613201*Z(18,19) +0.0254004*Z(32,31) -0.0017817*Z(48,49) 
+0.0001143*Z(70,69)

0.0039238 :  1:  +0.0631227*Z(2,3) +0.5235370*Z(8,7) -0.1974575*Z(16,17) +0.0351595*Z(30,29) 
-0.0029586*Z(46,47) +0.0001991*Z(68,67) -0.0000034*Z(92,93) +0.0000013*Z(122,121)

0.0020876 :  4:  +0.5264522*Z(14,15) -0.1989426*Z(26,25) +0.0395483*Z(40,41) -0.0039580*Z(60,59) 
+0.0003132*Z(82,83) -0.0000098*Z(110,109) +0.0000019*Z(140,141)

0.0015585 :  2:  +0.1070746*Z(6,5) +0.4864405*Z(12,13) -0.2575283*Z(24,23) +0.0619577*Z(38,39) 
-0.0077766*Z(58,57) +0.0006978*Z(80,81) -0.0000347*Z(108,107) +0.0000034*Z(138,139)

0.0015585 :  0:  +0.1070746*Z(1) +0.4864405*Z(4) -0.2575283*Z(11) +0.0619577*Z(22) -0.0077766*Z(37) 
+0.0006978*Z(56) -0.0000347*Z(79) +0.0000034*Z(106)

0.0010641 :  5:  +0.5130108*Z(20,21) -0.2283914*Z(34,33) +0.0540274*Z(50,51) -0.0069096*Z(72,71) 
+0.0006633*Z(96,97) -0.0000353*Z(126,125) +0.0000038*Z(158,159)

0.0007576 :  3:  +0.1466064*Z(10,9) +0.4417699*Z(18,19) -0.3046954*Z(32,31) +0.0926064*Z(48,49) 
-0.0154085*Z(70,69) +0.0017748*Z(94,95) -0.0001332*Z(124,123) +0.0000106*Z(156,157)

\end{verbatim}\normalsize

$\sigma_0=0.15$:
\small \begin{verbatim}
0.0544954 :  1:  +0.5580974*Z(2,3) -0.0824405*Z(8,7) +0.0063821*Z(16,17) -0.0000923*Z(30,29) 
+0.0000081*Z(46,47)

0.0132897 :  2:  +0.5500307*Z(6,5) -0.1247356*Z(12,13) +0.0147219*Z(24,23) -0.0006141*Z(38,39) 
+0.0000353*Z(58,57) +0.0000015*Z(80,81)

0.0132897 :  0:  +0.5500307*Z(1) -0.1247356*Z(4) +0.0147219*Z(11) -0.0006141*Z(22) +0.0000353*Z(37) 
+0.0000015*Z(56)

0.0045724 :  3:  +0.5384109*Z(10,9) -0.1664591*Z(18,19) +0.0266702*Z(32,31) -0.0019125*Z(48,49) 
+0.0001243*Z(70,69) +0.0000010*Z(124,123)

0.0035533 :  1:  +0.0787021*Z(2,3) +0.5165478*Z(8,7) -0.2092330*Z(16,17) +0.0387812*Z(30,29) 
-0.0034279*Z(46,47) +0.0002365*Z(68,67) -0.0000049*Z(92,93) +0.0000014*Z(122,121)

0.0020105 :  4:  +0.5253405*Z(14,15) -0.2016812*Z(26,25) +0.0404365*Z(40,41) -0.0040835*Z(60,59) 
+0.0003256*Z(82,83) -0.0000104*Z(110,109) +0.0000019*Z(140,141)

0.0014813 :  2:  +0.1160947*Z(6,5) +0.4804411*Z(12,13) -0.2640555*Z(24,23) +0.0649074*Z(38,39) 
-0.0083296*Z(58,57) +0.0007596*Z(80,81) -0.0000389*Z(108,107) +0.0000037*Z(138,139)

0.0014813 :  0:  +0.1160947*Z(1) +0.4804411*Z(4) -0.2640555*Z(11) +0.0649074*Z(22) -0.0083296*Z(37) 
+0.0007596*Z(56) -0.0000389*Z(79) +0.0000037*Z(106)

0.0010391 :  5:  +0.5122189*Z(20,21) -0.2300063*Z(34,33) +0.0546713*Z(50,51) -0.0070238*Z(72,71) 
+0.0006773*Z(96,97) -0.0000363*Z(126,125) +0.0000039*Z(158,159)

0.0007373 :  3:  +0.1509283*Z(10,9) +0.4377755*Z(18,19) -0.3077212*Z(32,31) +0.0945258*Z(48,49) 
-0.0158828*Z(70,69) +0.0018443*Z(94,95) -0.0001398*Z(124,123) +0.0000111*Z(156,157)

\end{verbatim}\normalsize

$\sigma_0=0.2$:
\small \begin{verbatim}
0.0384868 :  1:  +0.5555830*Z(2,3) -0.0978254*Z(8,7) +0.0082237*Z(16,17) -0.0001412*Z(30,29) 
+0.0000118*Z(46,47) +0.0000011*Z(68,67)

0.0112251 :  2:  +0.5475076*Z(6,5) -0.1351598*Z(12,13) +0.0166326*Z(24,23) -0.0007383*Z(38,39) 
+0.0000432*Z(58,57) +0.0000017*Z(80,81)

0.0112251 :  0:  +0.5475076*Z(1) -0.1351598*Z(4) +0.0166326*Z(11) -0.0007383*Z(22) +0.0000432*Z(37) 
+0.0000017*Z(56)

0.0041940 :  3:  +0.5364214*Z(10,9) -0.1725149*Z(18,19) +0.0282171*Z(32,31) -0.0020739*Z(48,49) 
+0.0001371*Z(70,69) +0.0000011*Z(124,123)

0.0032042 :  1:  +0.0929350*Z(2,3) +0.5092242*Z(8,7) -0.2203375*Z(16,17) +0.0424228*Z(30,29) 
-0.0039175*Z(46,47) +0.0002775*Z(68,67) -0.0000066*Z(92,93) +0.0000016*Z(122,121)

0.0019126 :  4:  +0.5238905*Z(14,15) -0.2051844*Z(26,25) +0.0415913*Z(40,41) -0.0042476*Z(60,59) 
+0.0003420*Z(82,83) -0.0000111*Z(110,109) +0.0000020*Z(140,141)

0.0013966 :  2:  +0.1252979*Z(6,5) +0.4738551*Z(12,13) -0.2708383*Z(24,23) +0.0681004*Z(38,39) 
-0.0089399*Z(58,57) +0.0008295*Z(80,81) -0.0000438*Z(108,107) +0.0000041*Z(138,139)

0.0013966 :  0:  +0.1252979*Z(1) +0.4738551*Z(4) -0.2708383*Z(11) +0.0681004*Z(22) -0.0089399*Z(37) 
+0.0008295*Z(56) -0.0000438*Z(79) +0.0000041*Z(106)

0.0010062 :  5:  +0.5111531*Z(20,21) -0.2321555*Z(34,33) +0.0555358*Z(50,51) -0.0071778*Z(72,71) 
+0.0006962*Z(96,97) -0.0000375*Z(126,125) +0.0000040*Z(158,159)

0.0007122 :  3:  +0.1559847*Z(10,9) +0.4329305*Z(18,19) -0.3112822*Z(32,31) +0.0968427*Z(48,49) 
-0.0164612*Z(70,69) +0.0019301*Z(94,95) -0.0001480*Z(124,123) +0.0000118*Z(156,157)

\end{verbatim}\normalsize

$\sigma_0=0.3$:
\small \begin{verbatim}
0.0207627 :  1:  +0.5499437*Z(2,3) -0.1254079*Z(8,7) +0.0120252*Z(16,17) -0.0002530*Z(30,29) 
+0.0000226*Z(46,47) +0.0000019*Z(68,67)

0.0078797 :  2:  +0.5419000*Z(6,5) -0.1556381*Z(12,13) +0.0207357*Z(24,23) -0.0010191*Z(38,39) 
+0.0000634*Z(58,57) +0.0000021*Z(80,81)

0.0078797 :  0:  +0.5419000*Z(1) -0.1556381*Z(4) +0.0207357*Z(11) -0.0010191*Z(22) +0.0000634*Z(37) 
+0.0000021*Z(56)

0.0034105 :  3:  +0.5317006*Z(10,9) -0.1859701*Z(18,19) +0.0318335*Z(32,31) -0.0024607*Z(48,49) 
+0.0001696*Z(70,69) -0.0000011*Z(94,95) +0.0000013*Z(124,123)

0.0025805 :  1:  +0.1179237*Z(2,3) +0.4939434*Z(8,7) -0.2406863*Z(16,17) +0.0496921*Z(30,29) 
-0.0049451*Z(46,47) +0.0003693*Z(68,67) -0.0000106*Z(92,93) +0.0000020*Z(122,121)

0.0016809 :  4:  +0.5202115*Z(14,15) -0.2137474*Z(26,25) +0.0444970*Z(40,41) -0.0046666*Z(60,59) 
+0.0003854*Z(82,83) -0.0000130*Z(110,109) +0.0000023*Z(140,141)

0.0012199 :  2:  +0.1430463*Z(6,5) +0.4596855*Z(12,13) -0.2843055*Z(24,23) +0.0748388*Z(38,39) 
-0.0102679*Z(58,57) +0.0009874*Z(80,81) -0.0000551*Z(108,107) +0.0000049*Z(138,139)

0.0012199 :  0:  +0.1430463*Z(1) +0.4596855*Z(4) -0.2843055*Z(11) +0.0748388*Z(22) -0.0102679*Z(37) 
+0.0009874*Z(56) -0.0000551*Z(79) +0.0000049*Z(106)

0.0009225 :  5:  +0.5083277*Z(20,21) -0.2377231*Z(34,33) +0.0578144*Z(50,51) -0.0075872*Z(72,71) 
+0.0007473*Z(96,97) -0.0000409*Z(126,125) +0.0000043*Z(158,159)

0.0006527 :  3:  +0.1670529*Z(10,9) +0.4216099*Z(18,19) -0.3191848*Z(32,31) +0.1022108*Z(48,49) 
-0.0178264*Z(70,69) +0.0021369*Z(94,95) -0.0001680*Z(124,123) +0.0000134*Z(156,157)

\end{verbatim}\normalsize

$\sigma_0=0.4$:
\small \begin{verbatim}
0.0120802 :  1:  +0.5438669*Z(2,3) -0.1492296*Z(8,7) +0.0157816*Z(16,17) -0.0003729*Z(30,29) 
+0.0000384*Z(46,47) +0.0000031*Z(68,67)

0.0055228 :  2:  +0.5358544*Z(6,5) -0.1747716*Z(12,13) +0.0249608*Z(24,23) -0.0013265*Z(38,39) 
+0.0000892*Z(58,57) +0.0000027*Z(80,81) +0.0000011*Z(108,107)

0.0055228 :  0:  +0.5358544*Z(1) -0.1747716*Z(4) +0.0249608*Z(11) -0.0013265*Z(22) +0.0000892*Z(37) 
+0.0000027*Z(56) +0.0000011*Z(79)

0.0027015 :  3:  +0.5263645*Z(10,9) -0.1998967*Z(18,19) +0.0358207*Z(32,31) -0.0029028*Z(48,49) 
+0.0002098*Z(70,69) -0.0000017*Z(94,95) +0.0000016*Z(124,123)

0.0020613 :  1:  +0.1388056*Z(2,3) +0.4783720*Z(8,7) -0.2587251*Z(16,17) +0.0568160*Z(30,29) 
-0.0060141*Z(46,47) +0.0004724*Z(68,67) -0.0000151*Z(92,93) +0.0000025*Z(122,121)

0.0014359 :  4:  +0.5158045*Z(14,15) -0.2234543*Z(26,25) +0.0479255*Z(40,41) -0.0051719*Z(60,59) 
+0.0004400*Z(82,83) -0.0000153*Z(110,109) +0.0000027*Z(140,141)

0.0010481 :  2:  +0.1591466*Z(6,5) +0.4448988*Z(12,13) -0.2970362*Z(24,23) +0.0817255*Z(38,39) 
-0.0116808*Z(58,57) +0.0011637*Z(80,81) -0.0000680*Z(108,107) +0.0000059*Z(138,139)

0.0010481 :  0:  +0.1591466*Z(1) +0.4448988*Z(4) -0.2970362*Z(11) +0.0817255*Z(22) -0.0116808*Z(37) 
+0.0011637*Z(56) -0.0000680*Z(79) +0.0000059*Z(106)

0.0008254 :  5:  +0.5047821*Z(20,21) -0.2444603*Z(34,33) +0.0606444*Z(50,51) -0.0081031*Z(72,71) 
+0.0008134*Z(96,97) -0.0000452*Z(126,125) +0.0000048*Z(158,159)

0.0005874 :  3:  +0.1782251*Z(10,9) +0.4090493*Z(18,19) -0.3273494*Z(32,31) +0.1081019*Z(48,49) 
-0.0193653*Z(70,69) +0.0023770*Z(94,95) -0.0001917*Z(124,123) +0.0000154*Z(156,157)

\end{verbatim}\normalsize

$\sigma_0=0.6$:
\small \begin{verbatim}
0.0048075 :  1:  +0.5316377*Z(2,3) -0.1875067*Z(8,7) +0.0226296*Z(16,17) -0.0006127*Z(30,29) 
+0.0000863*Z(46,47) +0.0000078*Z(68,67) +0.0000016*Z(92,93)

0.0028053 :  2:  +0.5234235*Z(6,5) -0.2079357*Z(12,13) +0.0331139*Z(24,23) -0.0019680*Z(38,39) 
+0.0001565*Z(58,57) +0.0000053*Z(80,81) +0.0000020*Z(108,107)

0.0028053 :  0:  +0.5234235*Z(1) -0.2079357*Z(4) +0.0331139*Z(11) -0.0019680*Z(22) +0.0001565*Z(37) 
+0.0000053*Z(56) +0.0000020*Z(79)

0.0016497 :  3:  +0.5149116*Z(10,9) -0.2263237*Z(18,19) +0.0440286*Z(32,31) -0.0038640*Z(48,49) 
+0.0003090*Z(70,69) -0.0000025*Z(94,95) +0.0000025*Z(124,123)

0.0013057 :  1:  +0.1704452*Z(2,3) +0.4481773*Z(8,7) -0.2887852*Z(16,17) +0.0702216*Z(30,29) 
-0.0081810*Z(46,47) +0.0007039*Z(68,67) -0.0000250*Z(92,93) +0.0000039*Z(122,121)

0.0009994 :  4:  +0.5058217*Z(14,15) -0.2435913*Z(26,25) +0.0554728*Z(40,41) -0.0063270*Z(60,59) 
+0.0005743*Z(82,83) -0.0000206*Z(110,109) +0.0000037*Z(140,141)

0.0007519 :  2:  +0.1855962*Z(6,5) +0.4154116*Z(12,13) -0.3194096*Z(24,23) +0.0951789*Z(38,39) 
-0.0146004*Z(58,57) +0.0015527*Z(80,81) -0.0000970*Z(108,107) +0.0000085*Z(138,139)

0.0007519 :  0:  +0.1855962*Z(1) +0.4154116*Z(4) -0.3194096*Z(11) +0.0951789*Z(22) -0.0146004*Z(37) 
+0.0015527*Z(56) -0.0000970*Z(79) +0.0000085*Z(106)

0.0006290 :  5:  +0.4963318*Z(20,21) -0.2595260*Z(34,33) +0.0672512*Z(50,51) -0.0093399*Z(72,71) 
+0.0009791*Z(96,97) -0.0000558*Z(126,125) +0.0000061*Z(158,159)

0.0004592 :  3:  +0.1984313*Z(10,9) +0.3826865*Z(18,19) -0.3427694*Z(32,31) +0.1202986*Z(48,49) 
-0.0226871*Z(70,69) +0.0029188*Z(94,95) -0.0002463*Z(124,123) +0.0000203*Z(156,157)

\end{verbatim}\normalsize

$\sigma_0=0.8$:
\small \begin{verbatim}
0.0022407 :  1:  +0.5203650*Z(2,3) -0.2161624*Z(8,7) +0.0283406*Z(16,17) -0.0008481*Z(30,29) 
+0.0001528*Z(46,47) +0.0000171*Z(68,67) +0.0000034*Z(92,93)

0.0015169 :  2:  +0.5114918*Z(6,5) -0.2346225*Z(12,13) +0.0403918*Z(24,23) -0.0025985*Z(38,39) 
+0.0002410*Z(58,57) +0.0000108*Z(80,81) +0.0000035*Z(108,107)

0.0015169 :  0:  +0.5114918*Z(1) -0.2346225*Z(4) +0.0403918*Z(11) -0.0025985*Z(22) +0.0002410*Z(37) 
+0.0000108*Z(56) +0.0000035*Z(79)

0.0010125 :  3:  +0.5034249*Z(10,9) -0.2493257*Z(18,19) +0.0518292*Z(32,31) -0.0048426*Z(48,49) 
+0.0004262*Z(70,69) -0.0000017*Z(94,95) +0.0000039*Z(124,123)

0.0008339 :  1:  +0.1919391*Z(2,3) +0.4207723*Z(8,7) -0.3123424*Z(16,17) +0.0821921*Z(30,29) 
-0.0102809*Z(46,47) +0.0009559*Z(68,67) -0.0000348*Z(92,93) +0.0000058*Z(122,121) 
+0.0000011*Z(154,155)

0.0006790 :  4:  +0.4953154*Z(14,15) -0.2625515*Z(26,25) +0.0631000*Z(40,41) -0.0075552*Z(60,59) 
+0.0007312*Z(82,83) -0.0000257*Z(110,109) +0.0000052*Z(140,141) +0.0000010*Z(176,175)

0.0005307 :  2:  +0.2050703*Z(6,5) +0.3877207*Z(12,13) -0.3376902*Z(24,23) +0.1076308*Z(38,39) 
-0.0174870*Z(58,57) +0.0019688*Z(80,81) -0.0001283*Z(108,107) +0.0000118*Z(138,139)

0.0005307 :  0:  +0.2050703*Z(1) +0.3877207*Z(4) -0.3376902*Z(11) +0.1076308*Z(22) -0.0174870*Z(37) 
+0.0019688*Z(56) -0.0001283*Z(79) +0.0000118*Z(106)

0.0004624 :  5:  +0.4870068*Z(20,21) -0.2747635*Z(34,33) +0.0743178*Z(50,51) -0.0107141*Z(72,71) 
+0.0011749*Z(96,97) -0.0000677*Z(126,125) +0.0000079*Z(158,159)

0.0003491 :  3:  +0.2146554*Z(10,9) +0.3567923*Z(18,19) -0.3560623*Z(32,31) +0.1321324*Z(48,49) 
-0.0260852*Z(70,69) +0.0035049*Z(94,95) -0.0003067*Z(124,123) +0.0000262*Z(156,157)

\end{verbatim}\normalsize

$\sigma_0=1.0$:
\small \begin{verbatim}
0.0011717 :  1:  +0.5104798*Z(2,3) -0.2379706*Z(8,7) +0.0329996*Z(16,17) -0.0010862*Z(30,29) 
+0.0002291*Z(46,47) +0.0000318*Z(68,67) +0.0000066*Z(92,93) +0.0000016*Z(122,121)

0.0008736 :  2:  +0.5006014*Z(6,5) -0.2559705*Z(12,13) +0.0466574*Z(24,23) -0.0031965*Z(38,39) 
+0.0003360*Z(58,57) +0.0000203*Z(80,81) +0.0000060*Z(108,107) +0.0000015*Z(138,139)

0.0008736 :  0:  +0.5006014*Z(1) -0.2559705*Z(4) +0.0466574*Z(11) -0.0031965*Z(22) +0.0003360*Z(37) 
+0.0000203*Z(56) +0.0000060*Z(79) +0.0000015*Z(106)

0.0006374 :  3:  +0.4925470*Z(10,9) -0.2687151*Z(18,19) +0.0588705*Z(32,31) -0.0057865*Z(48,49) 
+0.0005542*Z(70,69) +0.0000017*Z(94,95) +0.0000061*Z(124,123) +0.0000015*Z(156,157)

0.0005433 :  1:  +0.2065358*Z(2,3) +0.3967198*Z(8,7) -0.3309519*Z(16,17) +0.0926588*Z(30,29) 
-0.0122456*Z(46,47) +0.0012152*Z(68,67) -0.0000426*Z(92,93) +0.0000086*Z(122,121) 
+0.0000016*Z(154,155)

0.0004619 :  4:  +0.4849996*Z(14,15) -0.2794018*Z(26,25) +0.0703001*Z(40,41) -0.0087740*Z(60,59) 
+0.0009009*Z(82,83) -0.0000295*Z(110,109) +0.0000073*Z(140,141) +0.0000014*Z(176,175)

0.0003744 :  2:  +0.2190750*Z(6,5) +0.3626864*Z(12,13) -0.3524757*Z(24,23) +0.1188288*Z(38,39) 
-0.0202349*Z(58,57) +0.0023925*Z(80,81) -0.0001595*Z(108,107) +0.0000158*Z(138,139) 
+0.0000012*Z(174,173)

0.0003744 :  0:  +0.2190750*Z(1) +0.3626864*Z(4) -0.3524757*Z(11) +0.1188288*Z(22) -0.0202349*Z(37) 
+0.0023925*Z(56) -0.0001595*Z(79) +0.0000158*Z(106) +0.0000012*Z(137)

0.0003356 :  5:  +0.4775241*Z(20,21) -0.2890089*Z(34,33) +0.0812749*Z(50,51) -0.0121207*Z(72,71) 
+0.0013880*Z(96,97) -0.0000795*Z(126,125) +0.0000103*Z(158,159) +0.0000013*Z(196,195)

0.0002623 :  3:  -0.2270448*Z(10,9) -0.3325947*Z(18,19) +0.3671201*Z(32,31) -0.1431189*Z(48,49) 
+0.0293962*Z(70,69) -0.0041058*Z(94,95) +0.0003695*Z(124,123) -0.0000330*Z(156,157)

\end{verbatim}\normalsize

\section{Piston Removal}\label{app.pist}
The piston mode over the annulus is a mode without azimuth dependence, $q=0$,
where $K_p^{(0)}(x)$ is a constant that does not depend on the radial coordinate $x$.
Finite outer scales, $k_0>0$, prohibit the ultraviolet divergence of the
integrals that is characteristic to the Kolmogorov variance, so piston subtraction
could be implemented by ignoring the associated mode. Yet, removal of the
piston improves numerical stability because this contribution otherwise
dominates the matrix elements at $q=0$.

At $q=0$, equation (\ref{eq.Bail}) reads
\begin{eqnarray}
\frac{k}{2} J_0(kx) J_0(kx')
&=& \sum_{n\ge 0} (-1)^n(2n+1)J_{2n+1}(k)
\nonumber
\\
&& \times
\,_2F_1\left(\begin{array}{c}-n,n+1\\ 1\end{array}\mid \sin^2\phi\right)
\,_2F_1\left(\begin{array}{c}-n,n+1\\ 1\end{array}\mid \sin^2\Phi\right)
.
\label{eq.Bail0}
\end{eqnarray}
The dependence on $x$ and $x'$ is generated by the product of the $_2F_1(.)$,
which is a symmetric bi-variate polynomial in $x$ and $x'$ of mixed order $2n$.
The basic examples are
\begin{eqnarray}
&& \,_2F_1\left(\begin{array}{c}-n,n+1\\ 1\end{array}\mid \sin^2\phi\right)
\,_2F_1\left(\begin{array}{c}-n,n+1\\ 1\end{array}\mid \sin^2\Phi\right)
\nonumber \\
&&=
\left\{
\begin{array}{ll}
1 & $n=0$,\\
-1+2(x^2+x'^2) & $n=1$,\\
1-6(x^2+x'^2-x^4-x'^4-4x^2x'^2) & $n=2$,\\
-1+12(x^2+x'^2)-30(x^4+x'^4+4x^2x'^2)+20(x^6+x'^6+9x^4x'^2+9x^2x'^4) & $n=3$.\\
\end{array}
\right.
\label{eq.F2pist}
\end{eqnarray}
The simplicity of this format might be unexpected in view of the square roots in
(\ref{eq.Bailvars}), but if (\ref{eq.Bail}) is
constructed term-by-term on the right hand side by balancing the
powers in $k$ on both sides of the Neuman series, the linear coupling of $k$ and $x$, $x'$ in
the arguments on the left hand side makes this plausible.

The piston contribution is manipulated by adding some $\delta_n$ to
these products,
which by Mercer's theorem effects the piston's $\lambda_{p,0}^2$. To suppress the 
piston, equation (\ref{eq.evHybrid}) and the orthogonality
of the $K_p^{(0)}(x)$ suggest that
\begin{equation}
\int_\epsilon^1 x'dx' \int_\epsilon^1 x dx
\left[
\,_2F_1\left(\begin{array}{c}-n,n+1\\ 1\end{array}\mid \sin^2\phi\right)
\,_2F_1\left(\begin{array}{c}-n,n+1\\ 1\end{array}\mid \sin^2\Phi\right)
+\delta_n \right] =0.
\end{equation}
Insertion of (\ref{eq.F2pist}) yields for example
\begin{equation}
\delta_n =
\left\{
\begin{array}{ll}
-1 & $n=0$,\\
-1-2\epsilon^2 & $n=1$,\\
-5(1+2\epsilon^2+2\epsilon^4) & $n=2$,\\
-31(1+2\epsilon^2)-10\epsilon^4(8+7\epsilon^2) & $n=3$,\\
-229(1+2\epsilon^2)-2\epsilon^4(334+329\epsilon^2+294\epsilon^4) & $n=4$,\\
-1891(1+2\epsilon^2)-14\epsilon^4(403+488\epsilon^2+408\epsilon^4+396\epsilon^6) & $n=5$.\\
\end{array}
\right.
\end{equation}
The first value removes the entire term at $n=0$ in (\ref{eq.Bail0}).
The results in the main section added the $\delta_n$ to all $n\le 10$.

\bibliographystyle{apsrmp}
\bibliography{all}

\end{document}